\documentclass[12pt]{article}
\usepackage[papersize={8.5in,11in},textwidth=6in,bottom=1.25in]{geometry}

\usepackage{sectsty}
\sectionfont{\normalsize\bfseries}
\subsectionfont{\normalsize\sf}
\usepackage[mathcal]{euscript}
\usepackage[UKenglish]{babel}
\usepackage{bm}                     %AMS Packages
\usepackage{amsfonts}
\usepackage{amssymb}
\usepackage{amsmath}
\usepackage{amsthm}
\usepackage{graphicx}               %Graphics
\usepackage{natbib}                 %Citation style
\usepackage{colortbl}               %Tables
\usepackage{booktabs}
\usepackage{hyperref}
\usepackage{amssymb,amsmath}
\usepackage{multicol}

\usepackage[utf8]{inputenc}
\usepackage{graphicx,color}
\bibliography{referencias}

\newcommand{\abs}[1]{\ensuremath{|{#1}|}}

\usepackage{setspace}

\begin{document}
\begin{center}
\large \bf  The mRMR variable selection method: a comparative study for functional data\normalsize

\end{center}
\normalsize

\begin{center}
  Jos\'e R. Berrendero, Antonio Cuevas, Jos\'e L. Torrecilla \\
  Departamento de Matem\'aticas\\
  Universidad Aut\'onoma de Madrid, Spain
\end{center}

\begin{abstract}
	The use of variable selection methods is particularly appealing in statistical problems with functional data. The obvious general criterion for variable selection is to choose the `most representative' or `most relevant' variables. However, it is also clear that  a purely relevance-oriented criterion  could lead to select many redundant variables.   The mRMR (minimum Redundance Maximum Relevance) procedure, proposed by  Ding and Peng (2005) and Peng \it et al\rm. (2005) is an algorithm to systematically perform variable selection, achieving a reasonable trade-off between relevance and redundancy. In its original form, this procedure is based on the use of the so-called \it mutual information criterion\/ \rm to assess relevance and redundancy.  Keeping the focus on functional data problems, we propose here a modified version of the mRMR method, obtained by replacing the mutual information by the new association measure (called \it distance correlation\rm)  suggested by Sz\'ekely \it et al\rm. (2007). We  have also performed an extensive simulation study, including 1600 functional experiments (100 functional models $\times$ 4 sample sizes $\times$ 4 classifiers) and three real-data examples aimed at comparing the different versions of the mRMR methodology. The results are quite conclusive in favor of the new proposed alternative.
	

\end{abstract}

\section{Introduction}\label{sec:intro}

The use of high-dimensional or functional data entails some important practical issues. Besides the problems associated with computation time and storage costs, high-dimensionality introduces noise and redundancy. Thus, there is a strong case for using different techniques of dimensionality reduction. 

We will consider here dimensionality reduction via variable selection techniques. The general aim of these techniques is to replace the original high-dimensional (perhaps functional) data by lower dimensional projections obtained by just selecting a small subset of the original variables in each observation. In the 
case of functional data, this amounts to replace each observation $\{x(t),\ t\in[0,1]\}$ with a low-dimensional vector $(x(t_1),\ldots,x(t_k))$. Then, the chosen statistical methodology (supervised classification, clustering, regression,...) is performed with the `reduced', low-dimensional data. Usually the values $t_1,\ldots,t_k$ identifying the selected variables are the same for all considered data. A first advantage of variable selection (when compared with other dimension reduction methods, as Partial Least Squares) is the ease of interpretability, since the dimension reduction is made in terms of the original variables. In a way, variable selection appears as the most natural dimension reduction procedure in order to keep in touch, as much as possible, with the original data: see for instance \cite{gol99,lin09} among many other examples in experimental sciences or engineering. In \cite{gol99} the authors note that 50 genes (among almost 7000) are enough for cancer subtype classification. Likewise, \citet{lin09} point out that in some functional data regression (or classification) problems, as functional magnetic resonance imaging or gene expression, `the influence is concentrated at sensitive time points'.

We refer to \cite{guy06} for an account of different variable selection methods in the multivariate (non-functional) case. A partial comparative study, together with some new proposals for the functional framework, can be found in  \cite{ber14}.  

Throughout this work we will consider variable selection in the setting of functional supervised classification (the extension to more general regression problems is also possible with some obvious changes). Thus, the available sample information is a data set of type ${\mathcal D}_n=((X_1,Y_1),\ldots,(X_n,Y_n))$ of $n$ independent observations drawn from a random pair $(X,Y)$. Here $Y$ denotes a binary random variable, with values in $\{0,1\}$, indicating the membership to one of the populations $P_0$ or $P_1$ and $X_i$ are iid trajectories (in the space ${\mathcal C}[0,1]$ of real continuous functions on $[0,1]$), drawn from a stochastic process $X=X(t)$. The supervised classification  problem aims at predicting the membership class $Y$ of a new observation for which only the variable $X$ is known. Any function $g_n(x)=g_n(x;{\mathcal D}_n)$ with values in $\{0,1\}$ is called a classifier.

Several functional classifiers have been considered in the literature; see, e.g., \cite{bai11b} for a survey. Among them maybe the simplest one is the so-called $k$-nearest neighbours ($k$-NN) rule, according to which an observation $x$ is assigned to $P_1$ if 
and only if the majority among their $k$ nearest sample observations $X_i$ in the training sample fulfil $Y_i=1$. 
Here $k=k_n\in{\mathbb N}$ is a sequence of smoothing parameters which must satisfy $k_n\to \infty$ and $k_n/n\to 0$ in order to achieve consistency. In general, $k$-NN could be considered (from the limited experience so far available; see e.g., \cite{bai11a}) a sort of benchmark, reference method for functional supervised classification. Simplicity, ease of motivation and general good performance (it typically does not lead to gross classification errors) are perhaps the most attractive features of this method. Besides $k$-NN, we have also considered (inspired in the paper
by Ding and Peng \cite{din05} where a similar study is carried out) three additional classifiers: the popular \it Fisher's linear classifier \rm (LDA) used often in classical discriminant analysis, the so-called 
\it Na\"{\i}ve Bayes method \rm (NB) and the (linear) \it Support Vector Machine classifier \rm (SVM). 
Note that, in our empirical studies, all the mentioned classifiers ($k$-NN, LDA, NB and SVM) are used \it after the variable selection step\rm, on the `reduced data' resulting from the variable selection process.

In fact, as we will point out below, the main goal of our study is not to compare different classifiers. We are rather concerned with the comparison of different methods for variable selection (often referred to as \it feature selection\rm). A relevant procedure for variable selection, especially popular in the machine learning community, is the so-called \it minimum Redundancy Maximum Relevance \rm (mRMR) method. 
It was proposed by \citet{din05} and \citet{pen05}  as a tool 
to select the most discriminant subset of variables in the context of some relevant bioinformatics problems. See also \cite{bat94,kwa02,yu04} for closely related ideas.

\

\noindent \it The purpose of this paper\rm. 
Overall, we believe the mRMR procedure is a very natural way to tackle the variable selection problem if one wants to make completely explicit the trade-off relevance/redundancy. The method relies on the use of an association measure to assess the relevance and redundancy of the considered variables. In the original papers the so-called `mutual information' measure was used for this purpose.  The aim of the present paper is to propose other alternatives for the association measure, still keeping the main idea behind the mRMR procedure. 
In fact, most mRMR researchers admit that there is considerable room for improvement. We quote from
the discussion in \cite{pen05}: \it `The mRMR paradigm can be better
viewed as a general framework to effectively select features
and allow all possibilities for more sophisticated or more
powerful implementation schemes'\rm. 
In this vein, we consider several versions  of the mRMR and compare them by an extensive empirical study. Two of these versions are new: they are based on the `distance covariance' and `distance correlation' association measures proposed by 
\citet{sze07}. Our results suggest (and this is the main conclusion of our study) that the new version based on the distance correlation measure represents a clear improvement of the mRMR methodology.

The rest of the paper is organized as follows. Section \ref{sec:mrmr} contains a brief summary and some remarks about the mRMR algorithm. The different association measures under study (which are used to define the different versions of the  mRMR method) are explained in Section \ref{sec:measures}, with especial attention to the \textit{correlation of distances}. \cite{sze07,sze09} The empirical study, consisting of 1600 simulation experiments and some representative real data sets, is explained in Section \ref{sec:experiments}. Finally, some conclusions are given.

\section{The trade-off relevance/redundancy. The mRMR criterion}\label{sec:mrmr}

When faced with the problem of variable selection methods in high-dimensional (or functional) data sets, a natural idea arises at once: obviously, one should select the variables according to their relevance (representativeness). However, at the same time, one should avoid the redundancy which appears when two highly relevant variables  are closely associated to each other. In that case, one might expect that both variables essentially carry the same information, so that 
to choose just one of them should suffice.   

The mRMR variable selection method, as proposed in  \cite{din05,pen05}, provides a formal implementation of a variable selection procedure which explicitly takes into account  this trade-off relevance/redundancy. 

In our functional binary classification problem, the description of the mRMR method is as follows: the functional explanatory variable  $X(t)$, $t\in [0,1]$ will be used in a discretized version $(X(t_1),\ldots,X(t_N))$. When convenient, the notations $X_t$ and $X(t)$ will be used indistinctly. For any subset
$S$ of $\{t_1,\ldots,t_N\}$, the \it relevance\/ \rm and the \it redundancy\/ \rm of $S$ are defined, respectively, by 
\begin{equation}
\mbox{Rel}(S)=\frac{1}{\mbox{card}(S)}\sum_{t\in S}I(X_t,Y),\label{Rel}
\end{equation}
and
\begin{equation}
\mbox{Red}(S)=\frac{1}{\mbox{card}^2(S)}\sum_{s,t\in S}I(X_t,X_s),\label{Red}         
\end{equation}
where $\mbox{card}(S)$ denotes the cardinality of $S$ and $I(\cdot,\cdot)$ is an `association measure'. This function $I$ measures how much related are two variables. So, it is natural to think that the relevance of $X_t$ is measured by how much related it is with the response variable $Y$, that is $I(X_t,Y)$, whereas the redundancy between $X_t$ and $X_s$ is given by $I(X_s,X_t)$. Now, in summary, the mRMR algorithm  aims at maximizing the relevance avoiding an excess of redundancy. 

The choice of the association measure $I$ is a critical aspect in the mRMR methodology. In fact, this is the central point of the present work so that we will consider it in more detail later. 
By now, in order to explain how the mRMR method works, let us assume that the measure $I$ is given:

\begin{itemize}
	\item[(a)] The procedure starts by selecting the most relevant variable, given by the value $t_i$ 
	such that the set $S_i=\{t_i\}$ maximizes $\mbox{Rel}(S)$ among all the singleton sets of type $S_j=\{t_j\}$. 
	\item[(b)] Then, the variables are sequentially incorporated to the set $S$ of previously selected variables, with the criterion of maximizing the difference $\mbox{Rel}(S)-\mbox{Red}(S)$ (or alternatively the quotient $\mbox{Rel}(S)/\mbox{Red}(S)$).
	\item[(c)] Finally, different stopping rules can be considered.  We set the number of variables through a validation step (additional details can be found in Sections \ref{sec:experiments} and \ref{sec:reales}).
\end{itemize}

In practice, the use of the mRMR methodology is especially important in the functional data problems, where those variables which are very close together are often strongly associated. 

\begin{figure}
	\begin{center}
		\includegraphics[scale=0.50]{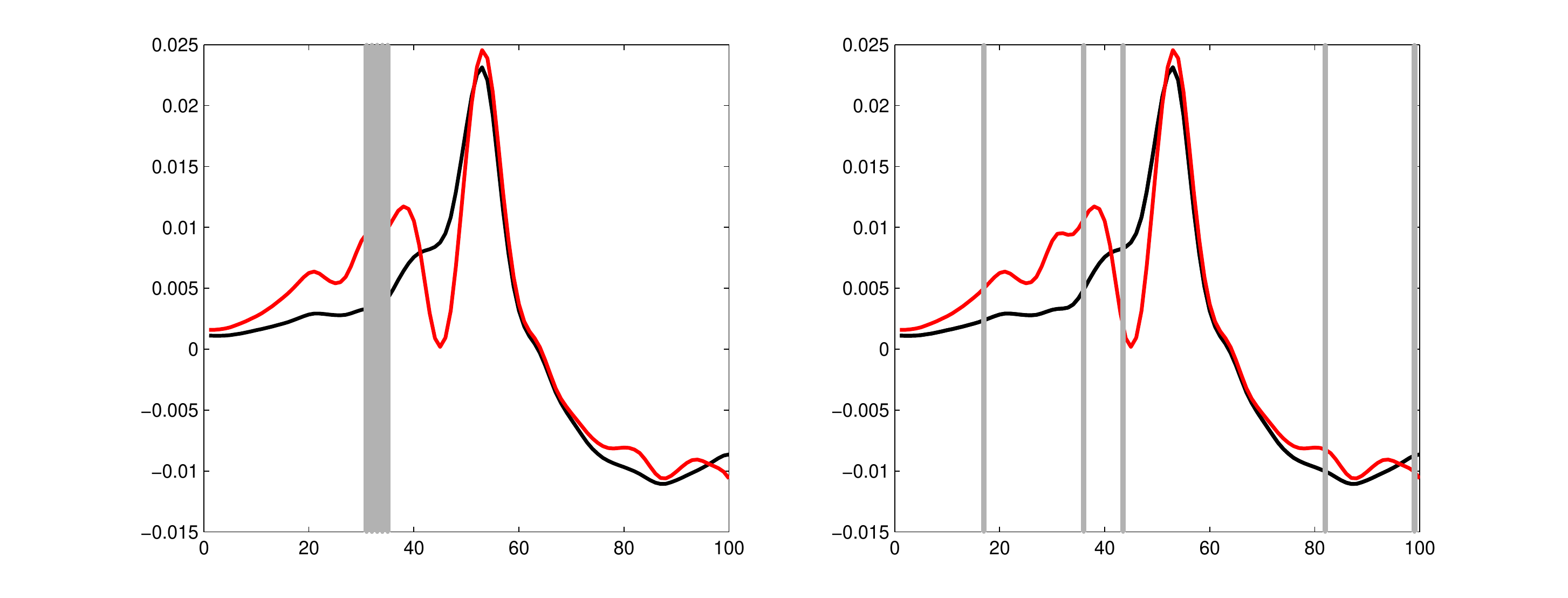}
		\caption{  Mean functions for both classes considered in the Tecator data set (first derivative). Left panel shows the five variables selected by Maximum Relevance. Right panel corresponds to the variables selected by mRMR. }
		\label{fig:mrmr}
	\end{center}
\end{figure}

The following example shows to what extent the mRMR makes a critical difference in the variable selection procedure. It concerns the well-known \it Tecator data set \rm (a benchmark example very popular in the literature on functional data; see Section \ref{sec:reales} for details). To be more specific, we use the first derivative of the curves in the Tecator data set, which is divided into two classes. We first use a simple `ranking procedure', where the variables are sequentially selected according to their relevance
(thus avoiding any notion of redundancy). The result is shown in the left panel of Figure \ref{fig:mrmr} (the selected variables are marked with grey vertical lines). It can be seen that in this case, all the five selected variables provide essentially the same information. On the right panel we see the variables selected from mRMR procedure which are clearly better placed to provide useful information. This visual impression is confirmed by comparing the error percentages obtained from a supervised classification method using only the variables selected by both methods. While the classification error obtained with the mRMR selected variables is 1.86\%, the corresponding error obtained with those of the ranking method is 4.09\%.   

\section{Association measures}\label{sec:measures}

As indicated in the previous section, the mRMR criterion relies on the use of an association measure $I(X,Y)$ between random variables. The choice of appropriate association measures is a classical issue in mathematical statistics. Many different proposals are available and, in several aspects, this topic is still open for further research, especially in connection with the use of high-dimensional data sets (arising, e.g., in genetic microarray examples,\cite{res11,hal11}).

A complete review of the main association measures for random variables is clearly beyond the scope of this paper. So, we will limit ourselves to present here the measures $I(X,Y)$ we have used in this work:

\begin{enumerate}
	\item \it The ordinary correlation coefficient between $X$ and $Y$ (in absolute value)\rm. This is the first obvious choice for the association measure $I(X,Y)$. It clearly presents some drawbacks (it does not characterize independence and it is unsuitable to
	capture non-linear association) but still, it does a good job in many practical situations. 
	\item The \it Mutual Information Measure\rm, $MI(X,Y)$  is defined by
	\begin{equation}
	MI(X,Y)=\int \log\frac{p(x,y)}{p_1(x)p_2(y)}p(x,y)d\mu(x,y),\label{MI}
	\end{equation} 
	where $X$, $Y$ are two random variables with respective $\mu$-densities $p_1$ and $p_2$; in the standard, absolutely continuous case, $\mu$ would be the product Lebesgue measure. In the discrete case, $\mu$ would be a counting measure on a countable support. The joint density of $(X,Y)$ is denoted by $p(x,y)$. 
	
	This is the association measure used in the original version of the mRMR procedure.\cite{din05,pen05}. 
	
	It is clear that $MI(X,Y)$ measures how far is $p(x,y)$ from the independence situation $p(x,y)=p_1(x)p_2(y)$. It is easily seen that $MI(X,Y)=MI(Y,X)$ and 
	$MI(X,Y)=0$ if and only if $X$ and $Y$ are independent. 
	
	In practice, $MI(X,Y)$ must be approximated by considering, if necessary, `discretized versions'  of $X$ and $Y$, obtained by grouping their values on intervals represented by suitable label marks, $a_i$, $b_j$. This leads to approximate expressions of type
	\begin{equation}
	\widehat {MI}(X,Y)=\sum_{i,j}\log\frac{{\mathbb P}(X=a_i,Y=b_j)}{{\mathbb P}(X=a_i){\mathbb P}(X=b_j)}{\mathbb P}(X=a_i,Y=b_j),\label{MIapp}
	\end{equation} 
	where, in turn, the probabilities can be empirically estimated by the corresponding relative frequencies.
	In \cite{din05} the authors suggest a threefold discretization pattern, i.e., the range of values of the variable is discretized in three classes. The limits of the discretization intervals are defined by the mean of the corresponding variable $\pm \sigma/2$ (where $\sigma$ is the standard deviation). We will explore this criterion in our empirical study below.    
	
	\item \it The Fisher-Correlation ($FC$) criterion\rm: It is a combination of the  $F$-statistic, 
	\begin{equation}
	F(X,Y)=\frac{\sum_k n_k(\bar{X}_k - \bar{X})^2/(K-1)}{\sum_k (n_k -1)\sigma_k^2/(n-K)},\label{F}
	\end{equation}
	used in the relevance measure (\ref{Rel}), and the ordinary correlation, $C$, used  in the redundancy measure (\ref{Red}). In the expression (\ref{F}), $K$ denotes the number of classes (so $K=2$ in our binary classification problem), 
	$\bar{X}$ denotes the mean of $X$, $\bar{X}_k$ is the mean value of $X$ of the elements belonging the $k$-th class, for $k=0,1$, and $n_k$ and $\sigma_k^2$ are the sample size and the variance of the $k$-th class, respectively.
	
	\citet{din05} suggest that, in principle, this criterion might look more useful than $\widehat {MI}$ when dealing with continuous variables but their empirical results do not support that idea. Such results are confirmed by our study so that, in general terms, we conclude that the mutual information (4) is a better choice even in the continuous setting. 
	
	\item \it Distance covariance\rm: this is an association measure recently proposed by \citet{sze07}. 
	Denote by $\varphi_{X,Y}$, $\varphi_{X}$, $\varphi_{Y}$    the characteristic functions
	of $(X,Y)$, $X$ and $Y$, respectively. Here $X$ and $Y$ denote multivariate random variables taking values in ${\mathbb R}^p$ and ${\mathbb R}^q$, respectively (note that the assumption $p=q$ is not needed). Let us suppose that the components of $X$ and $Y$ have finite first-order moments. 
	The distance covariance between $X$ and $Y$ is the non-negative value ${\cal V}(X,Y)$ defined by 
	\begin{equation} \label{dcov}
	{\cal V}^2(X,Y) = \int_ {\mathbb{R}^{p+q}} \mid \varphi_{X,Y}(u,v) -\varphi_X(u) \varphi_Y(v)\mid^2 w(u,v) du dv,
	\end{equation}
	with $w(u,v)= (c_p c_q \abs{u}_p^{1+p} \abs{v}_q^{1+q} )^{-1} $, where $c_d=\frac{\pi^{(1+d)/2}}{\Gamma((1+d)/2)}$ is half the surface area of the unit sphere in ${\mathbb R}^{d+1}$ and $|\cdot|_d$ stands for the Euclidean norm in ${\mathbb R}^d$. 
	
	While definition (\ref{dcov}) has a rather technical appearance, the resulting association measure has a number of interesting properties. Apart from the fact that (\ref{dcov}) allows for the case where $X$ and $Y$ have different dimensions, we have ${\cal V}^2(X,Y)=0$ if and only if $X$ and $Y$ are independent. Moreover, the indicated choice for the weights $w(u,v)$ provides valuable equivariance properties for ${\cal V}^2(X,Y)$ and the quantity can be consistently estimated from the mutual pairwise distances $|X_i-X_j|_p$ and $|Y_i-Y_j|_q$ between the sample values $X_i$ and $Y_j$ (no discretization is needed). 
	
	We refer to \cite{sze07,sze09,sze12,sze13} for a detailed study of this increasingly popular association measure. We refer also to \cite{ber14} for an alternative use (not related to mRMR) of ${\cal V}^2(X,Y)$ in variable selection.  
	
	\item \it Distance correlation\rm: this is just a sort of  standardized version of the distance covariance. If we denote ${\cal V}^2(X)={\cal V}^2(X,X)$, the (square) distance correlation between $X$ and $Y$ is
	defined by ${\cal R}^2(X,Y)=\frac{{\cal V}^2(X,Y)}{\sqrt{{\cal V}^2(X){\cal V}^2(Y)}}$ if ${\cal V}^2(X){\cal V}^2(Y)>0$, ${\cal R}^2(X,Y)$ $=0$ otherwise.

\end{enumerate}

Of course, other association measures might be considered. However, in order to get an affordable comparative study, we have limited our study to the main association measures previously used in the mRMR literature. We have only added the new measures ${\cal V}^2$ and ${\cal R}^2$, which we have tested as possible improvements of the method.

\

Also, alternative versions of the mRMR procedure have been proposed in literature. In particular, the Mutual Information measure could be estimated by kernel density estimation,\cite{wan95}. Regarding the kernel-based estimation of the MI measure, the crucial issue \cite{cao94} of the optimal selection of the smoothing parameter has not been, to our knowledge, explicitly addressed; note that here `optimal' should refer to the estimation of MI.
Likewise, other weighting factors might be used instead of just $card(S)$ in equation (2),\cite{est09}. However, still the `original' version of mRMR (with discretization-based MI estimation) seems to be the most popular standard; see \cite{man14,ngu14}  for  very recent examples.

Let us finally note that all the association measures we are considering take positive values. So, the phenomena associated with the the negative association values analyzed in \cite{dem13} do not apply in this case. 

\

\it Notation\rm. The association measures defined above will we denoted in the tables of our empirical study by \bf C\rm, \bf MI\rm, \bf FC\rm, \bf V\/ \rm and \bf R\rm, respectively.

\section{The simulation study}\label{sec:experiments}

We have checked five different versions of the mRMR variable selection methodology. They have been obtained by using different association measures (as indicated in the previous section) to assess relevance and redundancy. 

In all cases, the comparisons have been made in the context of problems of binary supervised classification, using 100 different models to generate the data $(X,Y)$. These models are defined either by 
\begin{itemize}
	\item[(i)] specifying the distributions of $X|Y=0$ and $X|Y=1$; in all cases, we take $p={\mathbb P}(Y=0)=1/2$.
	\item[(ii)] specifying both the marginal distribution of $X$ and the conditional distribution $\eta(x)={\mathbb P}(Y=1|X=x)$. 
\end{itemize}

Our experiments essentially consist of performing variable selection for each model using the different versions of mRMR and evaluating the results in terms of the respective probabilities of correct classification when different classifiers are used on the selected variables. The full list of considered models is available at 
the \textit{Supplemental material} document. All these models have been chosen in such a way that the optimal (Bayes) classification rule depends on just a finite number of variables. The processes considered include Brownian motion (with different mean functions), Brownian bridge  and several other Gaussian models, in particular the Ornstein-Uhlenbeck process. Other mixture models based on them are also considered. All these models are generated according to the pattern (i) above. In addition, we have considered several `logistic-type' models, generated by using pattern (ii). 

For each considered model all the variable selection methods (\bf C, \bf MI\rm, etc.) are checked for four sample sizes, $n=30,\ 50,\ 100,\ 200$ and four classification methods ($k$-\bf NN\rm, \bf LDA\rm, \bf NB\/ \rm and \bf SVM\rm). So, we have in total $100 \times  4 \times 4=1600$ simulation experiments.

\subsection{Classification methods}

We have used the four classifiers considered in the paper by Ding and Peng \cite{din05}, except that we have replaced the logistic regression classifier (which is closely related to the standard linear classifier) with the non-parametric $k$-NN method. All of them are widely known and details can be found, e.g. in \cite{has05}.

\begin{itemize}
	\item \textbf{Na\"{\i}ve Bayes classifier (NB).} This method relies on the assumption that the selected variables are Gaussian and conditionally independent in each class. So a new observation is assigned according to its posterior probability calculated from the Bayes rule. Of course the independence assumption will often fail (especially in the case of functional data). However, as shown in \cite{din05}, this rule works as an heuristics which offers sometimes a surprisingly good practical performance.  
	
	\item \textbf{The $k$-Nearest Neighbors classifier ($k$-\bf NN\rm).} According to this method (already commented in the introduction of the paper) a new observation is assigned to the class of the majority of its $k$ closest neighbors. We use the usual Euclidean distance (or $L^2$-distance when the method is used with the complete curves) to define the neighbors. The parameter $k$ is fitted through the validation step, as explained below.
	
	\item \textbf{Linear Discriminant Analysis (LDA).} The classic Fisher's linear discriminant is, still today, the most popular classification method among practitioners. It is know to be optimal under gaussianity and homoscedasticity of the distributions in both populations but, even when these conditions are not fulfilled, LDA tends to show a good practical performance in many real data sets. See, e.g., \cite{han06}. 
	
	\item \textbf{Support Vector Machine (SVM).} This is one of the most popular classification methodologies in  the last two decades. The basic idea is to look  for the `best hyperplane' in order to maximize the separation margin between the two classes. The use of different kernels (to send the observations to higher dimensional spaces where the separation is best achieved) is the most distinctive feature of this procedure. As in \cite{din05} we have used linear kernels.
	
\end{itemize}

As an objective reference, our simulation outputs include also the percentages of correct classification obtained with those classifiers  based on the complete curves, i.e., when no variable selection is done at all (except for LDA whose functional version is not feasible; see \cite{bai11b}). This reference method is called \bf Base\rm. A somewhat surprising conclusion of our study is that this \bf Base \rm method is often outperformed by the variable selection procedures. This could be due to the fact that the whole curves are globally more affected by noise than the selected variables. Thus, variable selection is beneficial not only in terms of simplicity but also in terms of accuracy.

\subsection{Computational details}

All codes have been implemented in MATLAB and are available from the authors upon request. We have used our own code for $k$-NN and LDA (which is a faster implementation of the MATLAB function \textit{classify}). The Na\"{\i}ve Bayes classifier is based on the MATLAB functions \textit{NaiveBayes.fit} and \textit{predict}. The linear SVM has been performed with the MATLAB version of the LIBLINEAR library (see \cite{fan08}) with bias and solver type 2, which obtains (with our data) very similar results to those of the default solver type 1 but faster.  The mRMR method has been implemented in such a way that different association measures can be used to define it. An online implementation of the original mRMR method can be found in \url{http://penglab.janelia.org/proj/mRMR/}  . 

Following \citet{din05}, the criteria (\ref{Rel}) and (\ref{Red}) to assess relevance and redundancy, respectively, are in fact replaced by approximate expressions, numbered (6) and (7) in \cite{din05}: as these authors point out, their expression (6) is equivalent to the relevance criterion (\ref{Rel}) while (7) provides an approximation for the minimum redundancy criterion (\ref{Red}).   The empirical estimation of the distance covariance (and distance correlation) implemented is the one proposed in \citet{sze07} expression (2.8).

All the functional simulated data  are \bf discretized \rm to $(x(t_1), \ldots, x(t_{100}))$, where $t_i$ are equi-spaced points in $[0,1]$. There is a partial exception in the case of the Brownian-like model, where (to avoid the degeneracy $x(t_0)=0$) we take $t_1=5/105$. Also (for a similar reason), a truncation is done at the end of the interval $[0,1]$ in those models including the Brownian Bridge.

%To simplify matters, the values ($0$ or $1$) of the membership variable $Y$ are not randomly assigned to the observations $X_i=X_i(t)$ in the simulation study (except in the case of the logistic-type models). Instead, for each considered simulated sample of size $n$, $n/2$ observation are drawn from each distribution $X|Y=i$, with $i=0,1$. 

The number $k$ of nearest neighbours in the $k$-NN classifier, the cost parameter $C$ of the SVM classifier and the number of selected variables are chosen by standard validation procedures.\cite{guy06}. To this end, in the simulation study, we have generated independent validation and test samples of size 200. Each simulation output is based on 200 independent runs.

\subsection{A few numerical outputs from the simulations}\label{outputs}

We present here just a small sample of the entire simulation outputs, which can be downloaded from \url{www.uam.es/antonio.cuevas/exp/mRMR-outputs.xlsx} . Some additional results, including a complete list of the considered models, can be found in the \it Supplemental material\/ \rm file. 

Tables \ref{tab:simnb} - \ref{tab:simsvm} contain the results obtained with NB, $k$-NN, LDA and SVM respectively. The  boxed outputs in these tables correspond to the winner and second best method in each row. The columns headings (MID, FCD, etc.) correspond to the different mRMR methods based on different association measures, as defined in Section \ref{sec:measures} (see the respective notations at the end of that section). The added letter `D' refers to the fact that global criterion to be maximized is just the difference between the measures (\ref{Rel}) and (\ref{Red}) of relevance and redundancy, respectively. There are other possibilities to combine (\ref{Rel}) and (\ref{Red}). One could take for instance the quotient. The corresponding outputs methods are denoted MIQ, FCQ, etc. in our supplementary material files. However, the outputs are not given here for the sake of brevity. In any case, our results suggest that the difference-based methods are globally (although not uniformly) better than those based on quotients.
The column `Base' gives the results  when no variable selection method is used (that is, the entire curves are considered). This column does not appear when the LDA method is used, since LDA cannot directly work on functional data. 

The row entries `Average accuracy' provide the average percentage of correct classification over the 100 considered model outputs; recall that every output is in turn obtained as an average over 200 independent runs. The rows `Average dim. red.' provide the average numbers of selected variables. The average number of times that every method beats the `Base' benchmark procedure is given in  `Victories over Base'.

\begin{table}
	\caption{Performance outputs for the  considered methods, using NB and the difference criterion, with different sample sizes. Each output is the result  of the 100 different models for each sample size.}{
		\footnotesize
		\begin{tabular}{llcccccc}\toprule      
			\small  Output (NB) &\small Sample size &\small MID &\small FCD &\small RD &\small VD &\small CD &\small Base\\
			\hline
			\small Average accuracy & $n=30$ & 78.08 & 78.42 & \framebox{79.56} & 79.24 & \framebox{79.28} & 77.28\\
			& $n=50$ & 79.64 & 79.34 & \framebox{80.92} & 80.45 & \framebox{80.46} & 78.29\\
			& $n=100$ & 80.76 & 80.06 & \framebox{81.90} & 81.34 & \framebox{81.41} & 78.84\\
			& $n=200$ & 81.46 & 80.44 & \framebox{82.55} & 81.90 & \framebox{82.05} & 79.13\\
			\hline
			\small Average dim. red & $n=30$ & 8.7 & 9.3 & \framebox{7.2} & \framebox{7.1} & 7.8 & 100\\
			& $n=50$ & 7.9 & 9.0 & \framebox{6.8} & \framebox{6.7} & 7.4 & 100\\
			& $n=100$ & 7.2 & 8.5 & \framebox{6.3} & \framebox{6.2} & 6.8 & 100\\
			& $n=200$ & 6.6 & 8.1 & \framebox{5.8} & \framebox{5.7} & 6.4 & 100\\
			\hline
			\small Victories over Base & $n=30$ & 57 & 61 & \framebox{77} & \framebox{71} & 69 & -\\
			& $n=50$ & 66 & 61 & \framebox{79} & \framebox{74} & 70 & -\\
			& $n=100$ & 77 & 61 & \framebox{88} & 81 & \framebox{85} & -\\
			& $n=200$ & 84 & 62 & \framebox{93} & 85 & \framebox{91} & -\\
			\hline
		\end{tabular}}\label{tab:simnb}
	\end{table}

	\begin{table}
		\caption{Performance outputs for the  considered methods, using $k$-NN and the difference criterion, with different sample sizes. Each output is the result  of the 100 different models for each sample size.}{
			\footnotesize
			\begin{tabular}{llcccccc}\toprule
				%&&\multicolumn{6}{c}{\small Methods} \\ \cline{3-8}
				\small Output ($k$-NN) & \small Sample size & \small MID & \small FCD & \small RD & \small VD & \small CD & \small Base\\
				\hline
				\small Avgerage accuracy & $n=30$ & 80.09 & 79.26 & \framebox{81.30} & \framebox{80.54} & 80.40 & 78.98\\
				& $n=50$ & 81.43 & 79.91 & \framebox{82.44} & \framebox{81.47} & 81.33 & 80.34\\
				& $n=100$ & \framebox{83.01} & 80.76 & \framebox{83.82} & 82.54 & 82.32 & 81.99\\
				& $n=200$ & \framebox{84.28} & 81.34 & \framebox{84.89} & 83.37 & 83.15 & 83.38\\
				\hline
				\small Average dim. red & $n=30$ & 9.2 & 9.8 & \framebox{7.7} & 8.3 & \framebox{8.0} & 100\\
				& $n=50$ & 9.3 & 9.9 & \framebox{7.9} & 8.5 & \framebox{8.1} & 100\\
				& $n=100$ & 9.6 & 10.2 & \framebox{8.2} & 8.7 & \framebox{8.3} & 100\\
				& $n=200$ & 9.8 & 10.4 & \framebox{8.5} & 8.8 & \framebox{8.7} & 100\\
				\hline
				\small Victories over Base & $n=30$ & 71 & 51 & \framebox{83} & \framebox{72} & 69 & -\\
				& $n=50$ & \framebox{71} & 45 & \framebox{81} & 70 & 68 & -\\
				& $n=100$ & \framebox{71} & 38 & \framebox{78} & 60 & 65 & -\\
				& $n=200$ & \framebox{73} & 33 & \framebox{82} & 56 & 58 & -\\
				\hline
			\end{tabular}}\label{tab:simknn}
		\end{table}

		\begin{table}
			\caption{Performance outputs for the considered methods, using LDA and the difference criterion, with different sample sizes. Each output is the result  of the 100 different models for each sample size.}{
				\footnotesize
				\begin{tabular}{llcccccc}\toprule    
					\small Output (LDA)&\small Sample size &\small MID &\small FCD &\small RD &\small VD &\small CD &\small Base\\
					\hline
					\small Avgerage accuracy & $n=30$ & \framebox{78.72} & 76.87 & \framebox{79.35} & 78.23 & 78.37 & -\\
					& $n=50$ & \framebox{80.28} & 77.84 & \framebox{80.59} & 79.15 & 79.36 & -\\
					& $n=100$ & \framebox{81.85} & 78.97 & \framebox{81.88} & 80.22 & 80.47 &-\\
					& $n=200$ & \framebox{82.96} & 79.83 & \framebox{82.87} & 81.02 & 81.30 & -\\
					\hline
					\small Average dim. red & $n=30$ & 5.6 & \framebox{4.9} & 5.0 & \framebox{4.6} & 5.2 & -\\
					& $n=50$ & 6.5 & 5.9 & \framebox{5.9} & \framebox{5.5} & 6.1 & -\\
					& $n=100$ & 7.9 & 7.5 & \framebox{7.1} & \framebox{6.8} & 7.4 & -\\
					& $n=200$ & 9.0 & 8.9 & \framebox{8.0} & \framebox{8.0} & 8.3 & -\\
					\hline
				\end{tabular}}\label{tab:simlda}
			\end{table}

			\begin{table}
				\caption{Performance outputs for the  considered methods, using SVM and the difference criterion, with different sample sizes. Each output is the result  of the 100 different models for each sample size.}{
					\footnotesize
					\begin{tabular}{llcccccc}\toprule
						\small Output (SVM) &\small Sample size &\small MID &\small FCD &\small RD &\small VD &\small CD &\small Base\\
						\hline
						\small Avgerage accuracy & $n=30$ & \framebox{81.53} & 79.41 & 81.50 & 80.35 & 80.51 & \framebox{81.91}\\
						& $n=50$ & \framebox{82.61} & 80.01 & 82.45 & 81.00 & 81.20 & \framebox{82.99}\\
						& $n=100$ & \framebox{83.75} & 80.75 & 83.45 & 81.77 & 82.00 & \framebox{84.11}\\
						& $n=200$ & \framebox{84.55} & 81.27 & 84.22 & 82.38 & 82.61 & \framebox{84.91}\\
						\hline
						\small Average dim. red & $n=30$ & 10.5 & 11.0 & \framebox{9.2} & 9.7 & \framebox{9.4} & 100\\
						& $n=50$ & 10.5 & 11.1 & \framebox{9.3} & 9.7 & \framebox{9.6} & 100\\
						& $n=100$ & 10.7 & 11.3 & \framebox{9.6} & 10.0 & \framebox{9.9} & 100\\
						& $n=200$ & 10.9 & 11.5 & \framebox{9.7} & 10.1 & \framebox{9.9} & 100\\
						\hline
						\small Victories over Base & $n=30$ & 37 & 39 & \framebox{49} & \framebox{43} & 42 & -\\
						& $n=50$ & 42 & 34 & \framebox{56} & 44 & \framebox{46} & -\\
						& $n=100$ & \framebox{49} & 32 & \framebox{57} & 41 & 47 & -\\
						& $n=200$ & 48 & 29 & \framebox{59} & 42 & \framebox{49} & -\\
						\hline
					\end{tabular}}\label{tab:simsvm}
				\end{table}
				
				It can be seen from these results that the global winner is the R-based mRMR method, with a especially good performance for small sample sizes. Note that the number of variables required by this method is also smaller, in general, than that of the remaining methods. Moreover, RD is the most frequent winner with respect to the Base method (with all classifiers) keeping, in addition, a more stable general performance when compared with the other variable selection methods. In this sense, R-based methods seem both efficient and reliable. In agreement with the results in \cite{din05}, the performance of the FC-based method is relatively poor. Finally, note that  the Base option (which uses the entire curves) is never the winner, with the partial exception of the SVM classifier. 
				
				\
				
				\subsection{Ranking the methods}\label{ranking} 
				
				It is not easy to draw general conclusions, and clear recommendations for practitioners, from a large simulation study. A natural idea is to  give some kind of quantitative assessment summarizing the relative merits of the different procedures. Many different ranking criteria might be considered. Following \citet{ber14}, we have considered here the following ones:
				\begin{itemize}
					\item \bf Relative ranking\rm: for each considered model and sample size the winner method (in terms of classification accuracy) gets 10 score points and the method with the worst performance gets 0 points.  The score 
					of any other method, with performance $u$, is defined by $10(u-w)/(W-w)$, where $W$ and $w$ denote, respectively, the performances of the best and the worst method. 
					\item \bf Positional ranking\rm: The winner gets 10 points, the second best gets 9, etc.
					\item \bf F1 ranking\rm: the scores are assigned according to the current criteria in a Formula 1 Grand Prix: the winner gets 25 score points and the following ones get 18, 15, 10, 8, 6, and 4 points. 
				\end{itemize}
				The summary results are shown in Tables \ref{tab:rknb} - \ref{tab:rksvm} and a visual version of the complete (400 experiments) relative ranking outputs for the $k$-NN classifier is displayed in Figure \ref{fig:rk} (analogous figures for the other classification methods can be found in the \textit{Supplemental material} document). The conclusions are self-explanatory and quite robust with respect to the ranking criterion. The mRMR methods based on the distance correlation measure are the uniform global winners. The results confirm the relative stability of R, especially when compared with MI whose good performance is restricted to a few models.
				
				Of course, the criteria for defining these rankings, as well as the idea of averaging over different models, are questionable (although one might think of a sort of Bayesian interpretation for these averages). Anyway, this is the only way we have found to provide an understandable summary for such a large  empirical study. On the other hand, since we have made available the whole outputs of our experiments, other different criteria 
				might be used by  interested readers.

				\begin{table}
					\caption{Global scores of the considered methods under three different ranking criteria using NB. Each output is the average of 100 models}{
						\footnotesize
						\begin{tabular}{llccccc}\toprule
							\small Ranking criterion (NB)& \small Sample size &\small MID & \small FCD & \small RD & \small VD & \small CD \\
							\hline
							\small Relative & $n=30$ & 2.43 & 5.10 & \framebox{8.67} & 7.08 & \framebox{8.10}\\
							& $n=50$ & 3.04 & 4.31 & \framebox{9.16} & 6.97 & \framebox{7.86}\\
							& $n=100$ & 3.38 & 3.92 & \framebox{9.28} & 6.84 & \framebox{7.82}\\
							& $n=200$ & 3.84 & 3.57 & \framebox{9.20} & 6.56 & \framebox{7.59}\\
							\hline
							\small Positional & $n=30$ & 6.65 & 7.62 & \framebox{8.84} & 8.21 & \framebox{8.68}\\
							& $n=50$ & 6.82 & 7.43 & \framebox{9.12} & 8.19 & \framebox{8.46}\\
							& $n=100$ & 6.87 & 7.36 & \framebox{9.26} & 8.16 & \framebox{8.35}\\
							& $n=200$ & 6.96 & 7.30 & \framebox{9.18} & 8.17 & \framebox{8.42}\\
							\hline
							\small F1 & $n=30$ & 11.64 & 15.11 & \framebox{18.64} & 16.37 & \framebox{18.24}\\
							& $n=50$ & 12.13 & 14.54 & \framebox{20.24} & 16.16 & \framebox{16.98}\\
							& $n=100$ & 12.19 & 14.29 & \framebox{20.82} & 16.17 & \framebox{16.53}\\
							& $n=200$ & 12.38 & 14.09 & \framebox{20.54} & 16.15 & \framebox{16.92}\\
							\hline
						\end{tabular}}\label{tab:rknb}
					\end{table}
					
					\begin{table}
						\caption{Global scores of the considered methods under three different ranking criteria using $k$-NN. Each output is the average of 100 models}{
							\footnotesize
							\begin{tabular}{llccccc}\toprule   
								\small Ranking criterion ($k$-NN) & \small Sample size &\small MID & \small FCD & \small RD & \small VD & \small CD \\
								\hline
								\small Relative & $n=30$ & 4.01 & 3.50 & \framebox{9.38} & 6.63 & \framebox{6.64}\\
								& $n=50$ & 4.66 & 3.09 & \framebox{9.07} & 6.19 & \framebox{6.34}\\
								& $n=100$ & 5.64 & 2.74 & \framebox{8.96} & \framebox{5.94} & 5.78\\
								& $n=200$ & \framebox{6.58} & 2.34 & \framebox{8.70} & 5.89 & 5.81\\
								\hline
								\small Positional & $n=30$ & 7.24 & 7.14 & \framebox{9.43} & \framebox{8.17} & 8.02\\
								& $n=50$ & 7.42 & 7.08 & \framebox{9.39} & \framebox{8.14} & 7.97\\
								& $n=100$ & 7.71 & 7.04 & \framebox{9.26} & \framebox{8.25} & 7.74\\
								& $n=200$ & 8.02 & 6.95 & \framebox{9.13} & \framebox{8.21} & 7.69\\
								\hline
								\small F1 & $n=30$ & 13.37 & 13.59 & \framebox{21.69} & \framebox{16.17} & 15.18\\
								& $n=50$ & 13.98 & 13.39 & \framebox{21.33} & \framebox{16.22} & 15.08\\
								& $n=100$ & 15.05 & 13.16 & \framebox{20.46} & \framebox{17.03} & 14.30\\
								& $n=200$ & 16.33 & 12.67 & \framebox{19.71} & \framebox{16.82} & 14.47\\
								\hline
							\end{tabular}}\label{tab:rkknn}
						\end{table}

						\begin{table}
							\caption{Global scores of the considered methods under three different ranking criteria using LDA. Each output is the average of 100 models}{
								\footnotesize
								\begin{tabular}{llccccc}\toprule
									\small Ranking criterion (LDA)& \small Sample size &\small MID & \small FCD & \small RD & \small VD & \small CD \\
									\hline
									\small Relative & $n=30$ & 5.00 & 1.98 & \framebox{8.94} & 6.24 & \framebox{6.47}\\
									& $n=50$ & 5.74 & 1.93 & \framebox{8.77} & 5.65 & \framebox{6.14}\\
									& $n=100$ & \framebox{6.07} & 1.94 & \framebox{8.51} & 5.50 & 5.95\\
									& $n=200$ & \framebox{6.53} & 2.08 & \framebox{8.44} & 5.36 & 5.92\\
									\hline
									\small Positional & $n=30$ & 7.57 & 6.68 & \framebox{9.31} & 8.17 & \framebox{8.27}\\
									& $n=50$ & 7.78 & 6.78 & \framebox{9.28} & 8.00 & \framebox{8.16}\\
									& $n=100$ & 7.85 & 6.90 & \framebox{9.14} & 8.02 & \framebox{8.09}\\
									& $n=200$ & 7.99 & 6.86 & \framebox{9.11} & 8.01 & \framebox{8.03}\\
									\hline
									\small F1 & $n=30$ & 14.69 & 11.81 & \framebox{20.86} & \framebox{16.51} & 16.13\\
									& $n=50$ & 15.56 & 12.13 & \framebox{20.60} & 15.72 & \framebox{15.99}\\
									& $n=100$ & 15.81 & 12.39 & \framebox{19.86} & \framebox{16.07} & 15.87\\
									& $n=200$ & \framebox{16.29} & 12.25 & \framebox{20.11} & 15.79 & 15.56\\
									\hline
								\end{tabular}}\label{tab:rklda}
							\end{table}

							\begin{table}
								\caption{Global scores of the considered methods under three different ranking criteria using SVM. Each output is the average of 100 models}{
									\footnotesize
									\begin{tabular}{llccccc}\toprule
										\small Ranking criterion (SVM)& \small Sample size &\small MID & \small FCD & \small RD & \small VD & \small CD \\
										\hline
										\small Relative & $n=30$ & \framebox{6.32} & 2.99 & \framebox{8.10} & 5.34 & 5.57\\
										& $n=50$ & \framebox{6.63} & 3 & \framebox{8.28} & 5.07 & 5.70\\
										& $n=100$ & \framebox{6.82} & 2.87 & \framebox{8.13} & 4.97 & 5.59\\
										& $n=200$ & \framebox{7.19} & 2.45 & \framebox{8.24} & 5.06 & 5.28\\
										\hline
										\small Positional & $n=30$ & \framebox{8.07} & 7.22 & \framebox{9.06} & 7.87 & 7.78\\
										& $n=50$ & \framebox{8.09} & 7.20 & \framebox{9.09} & 7.78 & 7.84\\
										& $n=100$ & \framebox{8.22} & 7.19 & \framebox{9.02} & 7.84 & 7.73\\
										& $n=200$ & \framebox{8.32} & 7.05 & \framebox{9.15} & 7.83 & 7.65\\
										\hline
										\small F1 & $n=30$ & \framebox{16.55} & 13.98 & \framebox{19.63} & 15.35 & 14.49\\
										& $n=50$ & \framebox{16.61} & 13.86 & \framebox{19.80} & 14.94 & 14.79\\
										& $n=100$ & \framebox{17.17} & 13.84 & \framebox{19.31} & 15.29 & 14.39\\
										& $n=200$ & \framebox{17.43} & 13.10 & \framebox{20.10} & 15.09 & 14.28\\
										\hline
									\end{tabular}}\label{tab:rksvm}
								\end{table}

								\begin{figure}
									\begin{center}
		\includegraphics[scale=0.30]{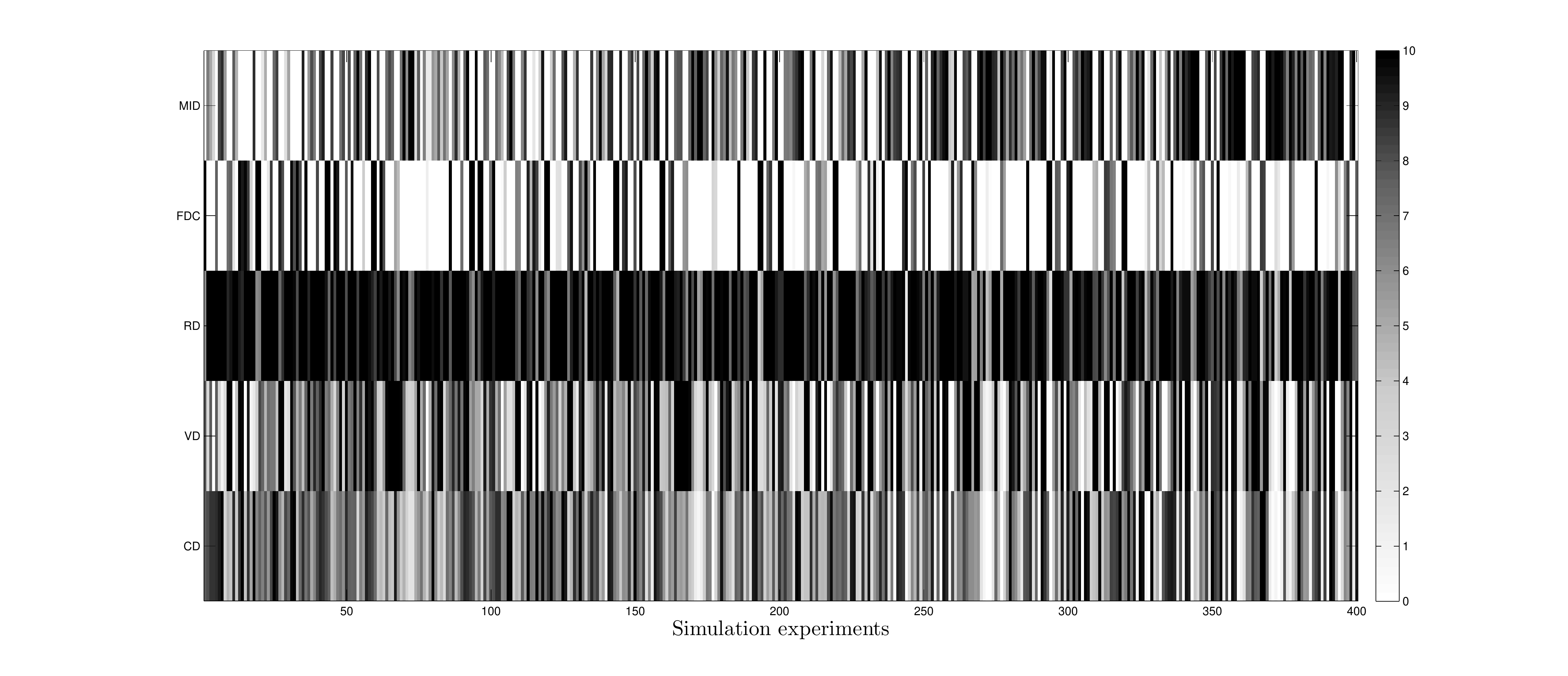}
										\caption{  Cromatic version of the global relative ranking table taking into account the 400 considered experiments (columns) and the difference-based mRMR versions  with the $k$-NN classifier: the darker de better. }
										\label{fig:rk}
									\end{center}
								\end{figure}

								\section{Real data examples}\label{sec:reales}
								
								We have chosen three real-data examples on the basis of their popularity in the literature on Functional Data Analysis: we call them \it Growth\/ \rm (93 growth curves in boys and girls), \it Tecator \/ \rm (215, near-infrared absorbance spectra from finely chopped meat) and \it Phoneme \/ \rm (1717 log-periodograms corresponding to the pronounciation of the sounds `aa' and `ao'). The respective dimensions of the considered discretizations for these  data are 31, 100 and 256. The second derivatives are used for the \it Tecator\/ \rm data. There are many references dealing with these data sets so we will omit here a detailed description of them. See, for example \citet{ram05}, \citet{fer06} and \citet{has05}, respectively, for additional details.  
								
								The methodology followed in the treatment of these data sets is similar to that followed in the simulation study, with a few technical differences. For \it Tecator\/ \rm and \it Growth\/ \rm data sets, a standard leave-one-out cross-validation is used. Such a procedure turns out to be too expensive (in computational terms) for the \it Phoneme\/ \rm data set. So in this case we have carried out 50-fold cross validation; see, for example, \cite[Sec. 7.10]{has05} for related ideas.

								A summary of the comparison outputs obtained for these data sets using the different mRMR criteria (as well as the  benchmark `Base' comparison, with no variable selection) is given in Table \ref{tab:real}. Again, the letter D in MID, FCD, etc. indicates that the relevance and redundancy measures are combined by difference. The analogous outputs using the quotient (instead of the difference) can be found in the \textit{Supplemental material} file.      
								
								\begin{table}
									\caption{Performances of the different mRMR methods in three data sets. From top to bottom tables stand for Naive Bayes, $k$-NN, LDA and linear SVM outputs respectively.}{
										\footnotesize
										\begin{tabular}{llcccccc}  \\
											\multicolumn{8}{c}{\small \bf NB outputs}\\
											\toprule
											\small Output & \small Data  &\small MID & \small FCD & \small RD & \small VD & \small CD &\small Base\\
											 \hline
											\small Classification accuracy & Growth & \framebox{92.47} & 87.10 & \framebox{89.25} & 87.10 & 86.02 & 84.95\\
											& Tecator & 98.60 & 97.67 & \framebox{99.53} & \framebox{99.53} & 98.14 & 97.21\\
											& Phoneme & 79.03 & \framebox{80.27} & \framebox{80.49} & 79.39 & 80.14 & 74.08\\
											\hline 
											\small Number of variables & Growth & 2.0 & \framebox{1.1} & 2.2 & \framebox{1.0} & 1.3 & 31\\
											& Tecator & 2.0 & 5.9 & \framebox{1.0} & \framebox{1.0} & 3.3 & 100\\
											& Phoneme & 12.6 & \framebox{10.3} & 15.8 & \framebox{5.8} & 15.9 & 256\\\hline
											
											\\  
											\multicolumn{8}{c}{\small \bf $k$-NN outputs}\\
											\toprule
											\small Output & \small Data  &\small MID & \small FCD & \small RD & \small VD & \small CD &\small Base\\
											\hline 
											\small  Classification accuracy & Growth & \framebox{95.70} & 83.87 & 94.62 & 91.40 & 84.95 & \framebox{96.77}\\
											& Tecator & 99.07 & 99.07 & \framebox{99.53} & \framebox{99.53} & 99.07 & 98.60\\
											& Phoneme & 80.14 & 80.48 & \framebox{81.14} & 80.31 & \framebox{80.55} & 78.80\\
											\hline 
											\small Number of variables & Growth & 3.5 & \framebox{1.0} & 2.5 & 4.8 & \framebox{1.1} & 31\\
											& Tecator & 5.7 & 3.0 & \framebox{1.0} & \framebox{1.0} & 4.0 & 100\\
											& Phoneme & 15.4 & \framebox{13.3} & 17.7 & 16.5 & \framebox{10.7} & 256\\\hline
											
											\\
											\multicolumn{8}{c}{\small \bf LDA outputs}\\
											\toprule
											\small Output & \small Data  &\small MID & \small FCD & \small RD & \small VD & \small CD &\small Base\\
											 \hline
											\small Classification accuracy & Growth & \framebox{94.62} & 91.40 & \framebox{94.62} & \framebox{94.62} & 89.25 & -\\
											& Tecator & \framebox{95.81} & 93.95 & 94.88 & \framebox{95.81} & 94.88 & -\\
											& Phoneme & \framebox{79.50} & 79.34 & 79.21 & 79.39 & \framebox{79.98} & -\\
											\hline 
											\small Number of variables & Growth & \framebox{3.4} & 5.0 & \framebox{3.1} & 4.2 & 5.0 & -\\
											& Tecator & \framebox{2.6} & 8.8 & 5.6 & \framebox{5.0} & \framebox{5.0} & -\\
											& Phoneme & 19.1 & \framebox{8.8} & 14.6 & 17.1 & \framebox{12.0} & -\\\hline
											
											\\
											\multicolumn{8}{c}{\small \bf SVM outputs}\\
											\toprule
											\small Output & \small Data  &\small MID & \small FCD & \small RD & \small VD & \small CD &\small Base\\
											\hline 
											\small Classification accuracy & Growth & 94.62 & 87.10 & 94.62 & \framebox{95.70} & 86.02 & \framebox{95.70}\\
											& Tecator & 98.14 & 99.07 & \framebox{99.53} & \framebox{99.53} & 98.60 & 99.07\\
											& Phoneme & \framebox{80.90} & 80.83 & 80.67 & 80.78 & 80.67 & \framebox{80.96}\\
											 \hline
											\small Number of variables & Growth & \framebox{3.4} & 5.0 & \framebox{2.5} & 4.2 & 5.0 & 31\\
											& Tecator & 6.7 & 2.0 & \framebox{1.0} & \framebox{1.0} & 4.1 & 100\\
											& Phoneme & 18.5 & \framebox{8.6} & 16.2 & 16.7 & \framebox{16.0} & 256\\
											\hline
										\end{tabular}}\label{tab:real}
									\end{table}

									The conclusions are perhaps less clear than those in the simulation study. The lack of a uniform winner is apparent.  However, the R-based method is clearly competitive and might even be considered as the global winner, taking into account both, accuracy and amount of dimension reduction. The \it Tecator\/ \rm outputs are particularly remarkable since RD and VD provide the best results (with three different classifiers) using just one variable. Again, variable selection methods beat here the `Base' approach (except for the Growth example) in spite of the drastic dimension reduction provided by the mRMR methods.

									\section{Final conclusions and comments}
									
									The mRMR methodology has become an immensely popular tool in the machine learning and bioinformatics communities.  For example, the  papers by \citet{din05} and \citet{pen05} had 819 and 2430 citations, respectively on Google Scholar (by October 2, 2014). As we have mentioned, these authors explicitly pointed out the need of further research, in order to get improved versions of the mRMR method. The idea would be to keep the basic mRMR paradigm but using other association measures (besides the mutual information). This paper exactly follows such line of research, with a particular focus on the classification problems involving functional data. 
									
									We think that the results are quite convincing: our extensive simulation study (based on 1600 simulation experiments and real data) places the mRMR method based in the R  association measure by \citet{sze07} globally above the original versions of the mRMR paradigm. This is perhaps the main conclusion of our work. The good performance of the distance correlation in comparison with the other measures can be partially explained by the facts that this measure captures non-linear dependencies (unlike C and FC), has a simple smoothing-free empirical estimator (dissimilar to MI) and is normalized (different from V).
									
									There are, however, some other more specific comments to be made.
									
									\begin{enumerate}
										
										\item  First of all, variable selection is worthwhile in functional data analysis. Accuracy can be kept (and often improved) using typically less than the 10\% of the original variables, with the usual benefits of the dimension reduction. This phenomenon happens for all the considered classifiers.
										
										\item The average number of selected variables with the R- or V-based methods is also smaller than that of MI and FC (that is, the standard mRMR procedures). This entails an interpretability gain: the fewer  selected variables, the stronger case for interpreting the meaning of such selection in the context of the considered problem. 
										
										\item The advantage of the R-based methods over the remaining procedures is more remarkable for the case of small sample sizes. This looks as a promising conclusion since small samples are very common in real problems (e.g. in biomedical research). 
										
										\item In those problems involving continuous variables there is a case for using non-parametric kernel density estimators in the empirical approximation of the mutual information criterion. However, these estimators are known to be highly sensitive to the selection of the smoothing parameter which can be seen as an additional unwelcome complication. On the other hand, the results reported so far (e.g. in \cite{pen05}) do not suggest that kernel estimators will lead to a substantial improvement over the simplest, much more popular  discretization estimators (see e.g. \cite{man14,ngu14}).
										
										\item Still in connection with the previous remark, it is worth noting the lack of smoothing parameters in the natural estimators of V and R.\cite{sze07} This can be seen as an additional advantage of the R- or V-based mRMR method.
										
										\item The better performance of R when compared with V can be explained by the fact that R is normalized so that relevance (\ref{Rel}) and redundancy (\ref{Red}) are always measured `in the same scale'. Otherwise, one of these two quantities could be overrated by the mRMR algorithm, specially when the difference criterion is used.

										\item The method FCD (sometimes suggested in the literature as a possible good choice) does not appear to be competitive. It is even defeated by the simple correlation-based method CD.
										
										\item In general, the difference-based methods are preferable to their quotient-based counterparts. The quotient-based procedures are only slightly preferable when combined with methods (FC, V) where relevance and redundancy are expressed in different scales.  The outputs for these quotient-based methods can be found in the complete list of results \url{www.uam.es/antonio.cuevas/exp/mRMR-outputs.xlsx}, and a summary is available in 
										\textit{Supplemental material} document.

										\item We should emphasize again that the goal of this paper is  to propose new  versions of the mRMR method and to compare them with the standard ones. Therefore, a wider study involving comparisons with other dimension reduction methods, is beyond the scope of this work. The recent paper by \citet{ber14} includes a study of this type (always in the functional setting) whose conclusions suggest that mRMR might be slightly outperformed  by the Maxima-Hunting (MH) procedure proposed by these authors. It also has a very similar performance to that of Partial Least Squares (PLS), although PLS is harder to interpret. Moreover, the number of variables selected by MH is typically smaller than those required by mRMR.

										\item Finally, if we had to choose just one among the considered classification methods, we should probably take $k$-NN. The above commented advantages in terms of ease of implementation and interpretability do not entail any significant price in efficiency. 
										
									\end{enumerate}

									\section*{Acknowledgements} This research has been partially supported by Spanish grant MTM2010-
									17366.

									\section*{Supplemental material}
									The \textit{Supplemental material} document contains: the complete list and description of all functional models, the summary Tables \ref{tab:simnb} - \ref{tab:real} with the quotient criterion instead of the difference one, figures analogous to Figure \ref{fig:rk} with NB, LDA and SVM, and some new tables with a few simulation results. All outputs (with both difference and quotient criteria) of the 1600 simulation experiments and real data can be found at \url{www.uam.es/antonio.cuevas/exp/mRMR-outputs.xlsx}.

\newpage
\begin{center}
	\large \bf  Supplementary material for the paper ``The mRMR variable selection method: a comparative study for functional data''
\end{center}
\normalsize

\begin{center}
	Jos\'e R. Berrendero, Antonio Cuevas, Jos\'e L. Torrecilla \\
	Departamento de Matem\'aticas\\
	Universidad Aut\'onoma de Madrid, Spain
\end{center}

\section{List of models used in the simulation study and a few example outputs}

Our simulation study consists of 400 experiments based on 100 different underlying models.  The optimal classification rule in each case depends only on a finite number of variables. 
Models differ in complexity and number of relevant variables. The processes involved are chosen among the following: first, the \bf standard Brownian Motion\rm, $B$. Second,
$BT$ denotes a \bf Brownian Motion with a trend \rm $m(t)$, i.e., $BT(t)$ $=B(t)+m(t)$; we have considered several choices for $m(t)$, a linear trend, $m(t)=ct$, a linear trend with random slope, i.e., $m(t)=\theta t$, where $\theta$ is a Gaussian r.v., and  different members of two  
parametric families: the \it peak\/ \rm functions
$\Phi_{m,k}$ and the \it hillside\/ \rm functions, defined by 
$$
\Phi_{m,k}=\int_0^t\varphi_{m,k}(s)ds \hspace{15pt} , \hspace{15pt} \mbox{hillside}_{t_0 ,b}(t)=b(t-t_0){\mathbb I}_{[t_0,\infty)},
$$
where, $\varphi_{m,k}(t)=\sqrt{2^{m-1}}\left[\mathbb{I}_{\left(\frac{2k-2}{2^m},\frac{2k-1}{2^m}\right)}-\mathbb{I}_{\left(\frac{2k-1}{2^m},\frac{2k}{2^m}\right)}\right]$ for $m\in\mathbb{N}$, 
$1\leq k\leq 2^{m-1}$.
Third, the \bf Brownian Bridge\rm: $BB(t)=B(t)-tB(1)$. Our fourth class of Gaussian processes is the \bf Ornstein–Uhlenbeck process\rm, with  zero mean ($OU$) or different mean functions $m(t)$ ($OUt$). Finally some ``smooth'' processes have been also include. They are obtained by convolving Brownian trajectories with Gaussian kernels. We have considered two levels of smoothing denoted by sB and ssB.

In the following list of models, $\mu_i$ denotes de distribution of $X|Y=i$ and \textit{variables} is the set of relevant variables in each Gaussian or Mixture case. We call them ``relevant'' in the sense that the optimal classification rule depends only on these variables. In the list below the variables written in boldface are ``especially relevant'' regarding their influence in the optimal classifier.

\

\noindent 1. \sc Gaussian models considered\rm:

\small

\

\begin{multicols}{2}
	\renewcommand{\labelenumi}{\scriptsize\arabic{enumi}.}
	\begin{enumerate}
		\item \textbf{G1} : $	\left \{ 
		\begin{matrix}
		\mu_0:&B(t)&\\
		\mu_1:&B(t)+\theta t&, \theta \sim N(0,3)
		\end{matrix}\right.$
		\item[] $variables=\{X_{100}\}$.
		
		\
		
		\item \textbf{G1b} : $	\left \{ 
		\begin{matrix}
		\mu_0:&B(t)&\\
		\mu_1:&B(t)+\theta t&, \theta \sim N(0,5)
		\end{matrix}\right.$
		\item[] $variables=\{X_{100}\}$.
		
		\
		
		\item \textbf{G2} : $	\left \{ 
		\begin{matrix}
		\mu_0:&B(t)+ t&\\
		\mu_1:&B(t)&
		\end{matrix}\right. $
		\item[] $variables=\{X_{100}\}$.
		
		\

		\item \textbf{G2b} : $	\left \{ 
		\begin{matrix}
		\mu_0:&B(t)+ 3t&\\
		\mu_1:&B(t)&
		\end{matrix}\right. $
		\item[] $variables=\{X_{100}\}$.
		
		\

		\item \textbf{G3} : $	\left \{ 
		\begin{matrix}
		\mu_0:&BB(t)\\
		\mu_1:&B(t)&
		\end{matrix}\right. $
		\item[] $variables=\{X_{100}\}$.
		
		\
		
		\item \textbf{G4} : $		\left \{ 
		\begin{matrix}
		\mu_0:&B(t)+hillside_{0.5,4}(t)&\\
		\mu_1:&B(t)&
		\end{matrix}\right.$
		\item[] $variables=\{X_{47},$\textbf{\textit{X}}$_{100}\}$.
		
		\
		
		\item \textbf{G5} : $		\left \{ 
		\begin{matrix}
		\mu_0:&B(t)+3\Phi_{1,1}(t)&\\
		\mu_1:&B(t)&
		\end{matrix}\right.$
		\item[] $variables=\{X_1,$\textbf{\textit{X}}$_{48},X_{100}\}$.
		
		\
		
		\item \textbf{G6} : $		\left \{ 
		\begin{matrix}
		\mu_0:&B(t)+5\Phi_{2,2}(t)&\\
		\mu_1:&B(t)&
		\end{matrix}\right.$
		\item[] $variables=\{X_{48},$\textbf{\textit{X}}$_{75},X_{100}\}$.
		
		\
		
		\item \textbf{G7} : $		\left \{ 
		\begin{matrix}
		\mu_0:&B(t)+5\Phi_{3,2}(t)+5\Phi_{3,4}(t)&\\
		\mu_1:&B(t)&
		\end{matrix}\right.$
		\item[] $variables=\{X_{22},$\textbf{\textit{X}}$_{35},X_{49},X_{74},$\textbf{\textit{X}}$_{88},X_{100}\}$.
		
		\
		
		\item \textbf{G8} :  $		\left \{ 
		\begin{matrix}
		\mu_0:&B(t)+3\Phi_{2,1.25}(t)+3\Phi_{2,2}(t)&\\
		\mu_1:&B(t)&
		\end{matrix}\right.$
		\item[] $variables=\{X_{9},$\textbf{\textit{X}}$_{35},X_{48},X_{62},$\textbf{\textit{X}}$_{75},X_{100}\}$.
		
	\end{enumerate}
	
\end{multicols}
\rm

\

\noindent 2. \sc logistic-type models considered\rm: they are all defined according standard (ii) (see Sec. 4 in the main paper). The process $X=X(t)$ follows one of the distributions mentioned above and $Y=\mbox{Binom}(1,\eta(X))$ with 
$\eta(x)=(1+e^{-\psi(x(t_1),\cdots,x(t_k))})^{-1}$,
a function of the relevant variables $x(t_1),\cdots,x(t_k)$. 

\

\textbf{L1:} $\psi(X)=10X_{65}$.

\

\textbf{L2:} $\psi(X)=10X_{30} + 10X_{70}$.

\

\textbf{L3:} $\psi(X)=10X_{30} - 10X_{70}$.

\

\textbf{L4:} $\psi(X)=20X_{30} + 50X_{50} 20X_{80}$.

\

\textbf{L5:} $\psi(X)=20X_{30} - 50X_{50}+ 20X_{80}$.

\

\textbf{L6:} $\psi(X)=10X_{10} + 30X_{40} + 10X_{72}+ 10X_{80} +20X_{95}$.

\

\textbf{L7:} $\psi(X)= \sum_{i=1}^{10} 10X_{10i}$.

\

\textbf{L8:} $\psi(X)= 20X_{30}^2 + 10X_{50}^4 + 50X_{80}^3$.

\

\textbf{L9:} $\psi(X)= 10X_{10} + 10|X_{50}| + 0X_{30}^2 X_{85}$.

\

\textbf{L10:} $\psi(X)= 20X_{33} + 20|X_{68}|$.

\

\textbf{L11:} $\psi(X)= \frac{20}{X_{35}}+\frac{30}{X_{77}}$.

\

\textbf{L12:} $\psi(X)= \log{X_{35} + \log{X_{77}}}$.

\

\textbf{L13:} $\psi(X)= 40X_{20}+30X_{28}+20X_{62}+10X_{67}$.

\

\textbf{L14:} $\psi(X)= 40X_{20}+30X_{28}-20X_{62}-10X_{67}$.

\

\textbf{L15:} $\psi(X)= 40X_{20}-30X_{28}+20X_{62}-10X_{67}$.

\

Some variations of these models have been also considered:

\

\textbf{L3b:} $\psi(X)=30X_{30} - 20X_{70}$.

\

\textbf{L4b:} $\psi(X)=30X_{30} + 20X_{50}+ 10X_{80}$.

\

\textbf{L5b:} $\psi(X)=10X_{30} - 10X_{50}+ 10X_{80}$.

\

\textbf{L6b:} $\psi(X)=20X_{10} + 20X_{40} + 20X_{72}+ 20X_{80} +20X_{95}$.

\

\textbf{L8b:} $\psi(X)=10X_{30}^2 + 10X_{50}^4+ 10X_{80}^3$.

\

\

\noindent 3. \sc Mixture-type models\rm: they are obtained by combining (via mixtures) in several ways the above mentioned Gaussian distributions assumed for $X|Y=0$ and $X|Y=1$. These models are denoted M1, ..., M10 in the output tables. 

\

\

\scriptsize
\begin{multicols}{2}
	\renewcommand{\labelenumi}{\scriptsize\arabic{enumi}.}

	\begin{enumerate}
		\item \textbf{M1} : $		\left \{ 
		\begin{matrix}
		\mu_0:&\left \{ 
		\begin{matrix}
		&B(t)+3t,& 1/2\\
		&B(t)-2t,& 1/2
		\end{matrix}\right.\\
		&&\\
		\mu_1:&B(t)& 
		\end{matrix}\right.$
		\item[] $variables=\{X_{100}\}$.
		
		\
		
		\item \textbf{M2} : $		\left \{ 
		\begin{matrix}
		\mu_0:&\left \{ 
		\begin{matrix}
		&B(t)+3\Phi_{2,2}(t),& 1/2\\
		&B(t)+5\Phi_{3,2}(t),& 1/2
		\end{matrix}\right.\\
		&&\\
		\mu_1:&B(t)& 
		\end{matrix}\right.$
		\item[] $variables=\{X_{22},$\textbf{\textit{X}}$_{35},X_{48},$\textbf{\textit{X}}$_{75},X_{100}\}$.
		
		\
		
		\item \textbf{M3} : $		\left \{ 
		\begin{matrix}
		\mu_0:&\left \{ 
		\begin{matrix}
		&B(t)+3\Phi_{2,2}(t),& 1/10\\
		&B(t)+5\Phi_{3,2}(t),& 9/10
		\end{matrix}\right.\\
		&&\\
		\mu_1:&B(t)& 
		\end{matrix}\right.$
		\item[] $variables=\{X_{22},$\textbf{\textit{X}}$_{35},X_{48},$\textbf{\textit{X}}$_{75},X_{100}\}$.

		\
		
		\item \textbf{M4}: $		\left \{ 
		\begin{matrix}
		\mu_0:&\left \{ 
		\begin{matrix}
		&B(t)+3\Phi_{2,2}(t),& 1/2\\
		&B(t)+5\Phi_{3,3}(t),& 1/2
		\end{matrix}\right.\\
		&&\\
		\mu_1:&B(t)& 
		\end{matrix}\right.$
		\item[] $variables=\{X_{48},$\textbf{\textit{X}}$_{62},$\textbf{\textit{X}}$_{75},X_{100}\}$.

		\
		
		\item \textbf{M5} :$		\left \{ 
		\begin{matrix}
		\mu_0:&\left \{ 
		\begin{matrix}
		&B(t)+3\Phi_{2,1}(t)&, 1/3\\
		&B(t)+3\Phi_{2,2}(t),& 1/3\\
		&B(t)+5\Phi_{3,2}(t),& 1/3
		\end{matrix}\right.\\
		&&\\
		\mu_1:&B(t)& 
		\end{matrix}\right.$
		\item[] $variables=\{X_{1},$\textbf{\textit{X}}$_{22},$\textbf{\textit{X}}$_{35},X_{48},$\textbf{\textit{X}}$_{75},X_{100}\}$.
		
		\
		
		\item \textbf{M6} : $		\left \{ 
		\begin{matrix}
		\mu_0:&\left \{ 
		\begin{matrix}
		&B(t)+3\Phi_{2,1}(t)&, 1/2\\
		&B(t)+3t&, 1/2
		\end{matrix}\right.\\
		&&\\
		\mu_1:&B(t)& 
		\end{matrix}\right.$
		\item[] $variables=\{X_{1},$\textbf{\textit{X}}$_{22},X_{49},$\textbf{\textit{X}}$_{100}\}$.

		\

		\item \textbf{M7} : $		\left \{ 
		\begin{matrix}
		\mu_0:&\left \{ 
		\begin{matrix}
		&B(t)+3\Phi_{1,1}(t)&, 1/2\\
		&BB(t)&, 1/2
		\end{matrix}\right.\\
		&&\\
		\mu_1:&B(t)& 
		\end{matrix}\right.$
		\item[] $variables=\{X_{1},$\textbf{\textit{X}}$_{48},$\textbf{\textit{X}}$_{100}\}$.
		
		\

		\item \textbf{M8} : $		\left \{ 
		\begin{matrix}
		\mu_0:&\left \{ 
		\begin{matrix}
		&B(t)+\theta t,\hspace{5pt} \theta \sim N(0,5)&, 1/2\\
		&B(t)+hillside_{0.5,5}(t)&, 1/2
		\end{matrix}\right.\\
		&&\\
		\mu_1:&B(t)& 
		\end{matrix}\right.$
		\item[] $variables=\{X_{47},$\textbf{\textit{X}}$_{100}\}$.

		\

		\item \textbf{M9} : $		\left \{ 
		\begin{matrix}
		\mu_0:&\left \{ 
		\begin{matrix}
		&B(t)+\theta t,\hspace{5pt} \theta \sim N(0,5)&, 1/2\\
		&BB(t)&, 1/2
		\end{matrix}\right.\\
		&&\\
		\mu_1:&B(t)& 
		\end{matrix}\right.$
		\item[] $variables=X_{100}$.

		\

		\item \textbf{M10} : $		\left \{ 
		\begin{matrix}
		\mu_0:&\left \{ 
		\begin{matrix}
		&B(t)+3\Phi_{1,1}(t)&, 1/3\\
		&B(t)-3t&, 1/3\\
		&BB(t)&, 1/3
		\end{matrix}\right.\\
		&&\\
		\mu_1:&B(t)& 
		\end{matrix}\right.$
		\item[] $variables=\{X_{1},$\textbf{\textit{X}}$_{48},$\textbf{\textit{X}}$_{100}\}$.
		
		\

		\item \textbf{M11} : $		\left \{ 
		\begin{matrix}
		\mu_0:&\left \{ 
		\begin{matrix}
		&B(t)+3\Phi_{1,1}(t)&, 1/4\\
		&B(t)-3t&, 1/4\\
		&B(t)+hillside_{0.5,5}(t)&, 1/4\\
		&BB(t)&, 1/4
		\end{matrix}\right.\\
		&&\\
		\mu_1:&B(t)& 
		\end{matrix}\right.$
		\item[] $variables=\{X_{1},$\textbf{\textit{X}}$_{48},$\textbf{\textit{X}}$_{100}\}$.
	\end{enumerate}
	\rm
\end{multicols}

\

\normalsize
Finally, the full list of models involved is
as follows:

\renewcommand{\labelenumi}{\scriptsize\arabic{enumi}.}
\footnotesize
\begin{multicols}{4}
	\columnseprule 0.25pt
	\begin{enumerate}
		\item L1 OU
		\item L1 OUt
		\item L1 B
		\item L1 sB
		\item L1 ssB
		\item L2 OU
		\item L2 OUt
		\item L2 B
		\item L2 sB
		\item L2 ssB
		\item L3 OU
		\item L3b OU
		\item L3 OUt
		\item L3b OUt
		\item L3 B
		\item L3b B
		\item L3 sB
		\item L3 ssB
		\item L4 OU
		\item L4b OU
		\item L4 OUt
		\item L4b OUt
		\item L4 B
		\item L4 sB
		\item L4 ssB
		\item L5 OU
		\item L5b OU
		\item L5 OUt
		\item L5 B
		\item L5 sB
		\item L5 ssB
		\item L6 OU
		\item L6b OU
		\item L6 OUt
		\item L6b OUt
		\item L6 B
		\item L6 sB
		\item L6 ssB
		\item L7 OU
		\item L7b OU
		\item L7 OUt
		\item L7b OUt
		\item L7 B
		\item L7 sB
		\item L7 ssB
		\item L8 B
		\item L8 sB
		\item L8 ssB
		\item L8b OU
		\item L9 B
		\item L9 sB
		\item L9 ssB
		\item L10 OU
		\item L10 B
		\item L10 sB
		\item L10 ssB
		\item L11 OU
		\item L11 OUt
		\item L11 B
		\item L11 sB
		\item L11 ssB
		\item L12 OU
		\item L12 OUt
		\item L12 B
		\item L12 sB
		\item L12 ssB
		\item L13 OU
		\item L13 OUt
		\item L13 B
		\item L13 sB
		\item L13 ssB
		\item L14 OU
		\item L14 OUt
		\item L14 B
		\item L14 sB
		\item L15 OU
		\item L15 OUt
		\item L15 B
		\item L15 sB
		\item G1
		\item G1b
		\item G2
		\item G2b
		\item G3
		\item G4
		\item G5
		\item G6
		\item G7
		\item G8
		\item M1
		\item M2
		\item M3
		\item M4
		\item M5
		\item M6
		\item M7
		\item M8
		\item M9
		\item M10
		\item M11
	\end{enumerate}
\end{multicols}

\section*{Simulation results}

We next provide a few simulation results. See \url{www.uam.es/antonio.cuevas/exp/mRMR-outputs.xlsx} for the full simulation outputs. 

\

\begin{center}\footnotesize
	\begin{tabular}{lcccccc}
		\multicolumn{7}{c}{\bf \small NB accuracy outputs}\\
		\hline\noalign{\smallskip}
		Model & MID & FCD & RD & VD & CD & Base\\
		\hline\noalign{\smallskip}
		L7\_OU & \framebox{90.73} & 87.01 & 90.41 & 87.72 & 90.50 & \framebox{93.35}\\
		L1\_OUt & 72.54 & \framebox{74.92} & 74.10 & \framebox{74.56} & 74.26 & 73.33\\
		L14\_B & 75.89 & \framebox{77.31} & \framebox{77.17} & 76.72 & 76.75 & 75.33\\
		L9\_sB & 84.73 & \framebox{86.01} & \framebox{85.84} & 85.40 & 85.78 & 82.27\\
		G1b & 79.74 & 75.94 & \framebox{80.73} & \framebox{81.66} & 77.67 & 79.49\\
		G3 & 76.22 & 64.54 & \framebox{78.84} & \framebox{78.78} & 72.67 & 71.24\\
		G6 & 83.13 & 83.50 & \framebox{84.38} & 83.96 & \framebox{84.28} & 74.90\\
		M1 & 79.12 & 73.09 & \framebox{80.93} & \framebox{81.89} & 76.77 & 77.61\\
		M4 & 64.73 & \framebox{68.04} & \framebox{67.91} & 67.48 & 67.89 & 61.36\\
		M6 & 81.25 & 80.09 & \framebox{83.06} & 82.45 & \framebox{83.21} & 78.02\\
		\noalign{\smallskip}\hline
	\end{tabular} 
\end{center}   
\noindent\footnotesize Table 10.- Average NB accuracy (proportion of correct classification) outputs, over 200 runs of the considered methods with sample size $n=50$.\normalsize

\begin{center}\footnotesize
	\begin{tabular}{lcccccc}
		\multicolumn{7}{c}{\bf \small NB number of variables}\\
		\hline\noalign{\smallskip}
		Model & MID & FCD & RD & VD & CD & Base\\
		\hline\noalign{\smallskip}
		L7\_OU & 12.1 & 14.1 & \framebox{10.1} & 11.5 & \framebox{11.1} & 100\\
		L1\_OUt & 8.9 & 8.0 & 7.0 & \framebox{5.9} & \framebox{6.6} & 100\\
		L14\_B & 7.9 & 8.0 & \framebox{6.0} & 7.5 & \framebox{5.9} & 100\\
		L9\_sB & \framebox{4.9} & 7.5 & \framebox{4.7} & 4.9 & 5.3 & 100\\
		G1b & 6.9 & 13.4 & \framebox{4.8} & \framebox{1.5} & 9.0 & 100\\
		G3 & 6.4 & 11.6 & \framebox{3.7} & \framebox{3.7} & 8.2 & 100\\
		G6 & 3.4 & \framebox{1.9} & 3.5 & 3.1 & \framebox{2.7} & 100\\
		M1 & 4.9 & 11.7 & \framebox{4.5} & \framebox{1.4} & 8.4 & 100\\
		M4 & 6.5 & 5.0 & 5.1 & \framebox{3.4} & \framebox{4.6} & 100\\
		M6 & 7.1 & 7.6 & \framebox{5.5} & \framebox{5.6} & 6.4 & 100\\      
		\noalign{\smallskip}\hline 
	\end{tabular} 
\end{center}   
\noindent\footnotesize Table 11.- Average number of selected variables over 200 runs of the considered methods with sample size $n=50$ using  NB.\normalsize

\newpage

\begin{center}\footnotesize
	\begin{tabular}{lcccccc}
		\multicolumn{7}{c}{\bf \small $k$-NN accuracy outputs}\\
		\hline\noalign{\smallskip}
		Model & MID & FCD & RD & VD & CD & Base\\
		\hline\noalign{\smallskip}
		L7\_OU & 90.74 & 86.89 & \framebox{90.78} & 87.79 & 90.67 & \framebox{92.21}\\
		L1\_OUt & 75.83 & \framebox{77.34} & 76.77 & \framebox{77.22} & 77.11 & 75.81\\
		L14\_B & 75.34 & \framebox{77.16} & \framebox{77.29} & 76.20 & 76.49 & 74.43\\
		L9\_sB & 86.79 & \framebox{87.39} & \framebox{87.35} & 87.03 & 87.28 & 86.10\\
		G1b & 79.11 & 75.74 & \framebox{80.00} & \framebox{80.01} & 78.13 & 78.57\\
		G3 & 73.39 & 61.97 & \framebox{77.28} & \framebox{77.03} & 68.20 & 65.26\\
		G6 & \framebox{91.95} & 84.16 & 88.22 & 85.80 & 84.68 & \framebox{92.19}\\
		M1 & 81.63 & 75.47 & \framebox{82.79} & \framebox{83.00} & 80.59 & 80.72\\
		M4 & 72.94 & 71.60 & \framebox{74.72} & 70.30 & 71.66 & \framebox{73.29}\\
		M6 & 83.08 & 79.70 & \framebox{84.13} & \framebox{84.26} & 84.02 & 80.99\\
		\noalign{\smallskip}\hline 
	\end{tabular} 
\end{center}   
\noindent\footnotesize Table 12.- Average  $k$-NN accuracy (proportion of correct classification) outputs, over 200 runs of the considered methods with sample size $n=50$. \normalsize

\

\

\begin{center}\footnotesize
	\begin{tabular}{lcccccc}
		\multicolumn{7}{c}{\bf \small $k$-NN number of variables}\\
		\hline\noalign{\smallskip}
		Model & MID & FCD & RD & VD & CD & Base\\
		\hline\noalign{\smallskip}
		L7\_OU & 11.5 & 14.5 & \framebox{10.4} & 12.1 & \framebox{11.3} & 100\\
		L1\_OUt & 9.0 & 7.9 & 6.9 & \framebox{6.5} & \framebox{6.8} & 100\\
		L14\_B & 8.3 & 7.6 & \framebox{5.5} & 8.0 & \framebox{6.5} & 100\\
		L9\_sB & 6.3 & 7.7 & \framebox{6.0} & 7.1 & \framebox{6.0} & 100\\
		G1b & 7.8 & 11.7 & \framebox{6.5} & \framebox{6.3} & 8.7 & 100\\
		G3 & 5.1 & 11.2 & \framebox{2.5} & \framebox{2.9} & 7.8 & 100\\
		G6 & 11.5 & 12.6 & 9.0 & \framebox{8.2} & \framebox{7.5} & 100\\
		M1 & 7.8 & 11.4 & \framebox{6.3} & \framebox{4.9} & 8.6 & 100\\
		M4 & 11.7 & 16.1 & 10.4 & \framebox{9.7} & \framebox{10.1} & 100\\
		M6 & 9.9 & 9.6 & \framebox{7.5} & 8.7 & \framebox{7.6} & 100\\
		\noalign{\smallskip}\hline 
	\end{tabular} 
\end{center}   
\noindent\footnotesize Table 13.- Average number of selected variables over 200 runs of the considered methods with sample size $n=50$ using  $k$-NN.\normalsize

\newpage

\begin{center}\footnotesize
	\begin{tabular}{lcccccc}
		\multicolumn{7}{c}{\bf \small LDA accuracy outputs}\\
		\hline\noalign{\smallskip}
		Model & MID & FCD & RD & VD & CD & Base\\
		\hline\noalign{\smallskip}
		L7\_OU & 89.27 & 85.13 & \framebox{89.75} & 86.81 & \framebox{90.10} & -\\
		L1\_OUt & 72.05 & \framebox{73.74} & 73.48 & \framebox{73.96} & 73.66 & -\\
		L14\_B & 75.25 & \framebox{76.35} & \framebox{77.12} & 75.62 & 76.27 & -\\
		L9\_sB & 84.91 & 84.88 & \framebox{84.96} & 84.69 & \framebox{85.12} & -\\
		G1b & 53.35 & 52.22 & \framebox{54.15} & \framebox{54.49} & 51.67 & -\\
		G3 & 52.26 & 50.91 & \framebox{53.53} & \framebox{53.51} & 51.06 & -\\
		G6 & \framebox{95.28} & 87.92 & \framebox{90.54} & 86.59 & 87.80 & -\\
		M1 & 54.44 & 53.64 & \framebox{54.88} & \framebox{54.68} & 53.70 & -\\
		M4 & \framebox{78.95} & 70.98 & \framebox{76.46} & 71.37 & 71.30 & -\\
		M6 & 79.92 & 79.53 & \framebox{80.80} & 80.57 & \framebox{80.70} & -\\
		\noalign{\smallskip}\hline 
	\end{tabular} 
\end{center}   
\noindent\footnotesize Table 14.- Average LDA accuracy (proportion of correct classification) outputs, over 200 runs of the considered methods with sample size $n=50$.\normalsize

\

\

\begin{center}\footnotesize
	\begin{tabular}{lcccccc}
		\multicolumn{7}{c}{\bf \small LDA number of variables}\\
		\hline\noalign{\smallskip}
		Model & MID & FCD & RD & VD & CD & Base\\
		\hline\noalign{\smallskip}
		L7\_OU & \framebox{5.5} & 6.9 & 6.2 & \framebox{6.1} & 7.2 & -\\
		L1\_OUt & 5.6 & 4.6 & \framebox{4.6} & \framebox{4.4} & 4.9 & -\\
		L14\_B & 3.6 & \framebox{3.2} & \framebox{3.1} & 4.1 & 3.6 & -\\
		L9\_sB & 4.0 & 3.5 & \framebox{3.3} & \framebox{3.4} & 3.5 & -\\
		G1b & 5.6 & 7.0 & \framebox{5.3} & \framebox{4.6} & 7.4 & -\\
		G3 & 7.1 & 8.9 & \framebox{5.0} & \framebox{5.0} & 8.9 & -\\
		G6 & \framebox{10.7} & 14.3 & 11.2 & \framebox{7.8} & 11.6 & -\\
		M1 & \framebox{5.5} & 7.2 & \framebox{5.3} & 5.7 & 6.8 & -\\
		M4 & 11.1 & 10.2 & 10.3 & \framebox{7.4} & \framebox{8.5} & -\\
		M6 & 6.4 & \framebox{4.5} & 5.2 & 5.2 & \framebox{4.8} & -\\    
		\noalign{\smallskip}\hline 
	\end{tabular} 
\end{center}   
\noindent\footnotesize Table 15.- Average number of selected variables over 200 runs of the considered methods with sample size $n=50$ using  LDA.\normalsize

\newpage

\begin{center}\footnotesize
	\begin{tabular}{lcccccc}
		\multicolumn{7}{c}{\bf \small SVM accuracy outputs}\\
		\hline\noalign{\smallskip}
		Model & MID & FCD & RD & VD & CD & Base\\
		\hline\noalign{\smallskip}
		L7\_OU & \framebox{90.73} & 87.01 & 90.41 & 87.72 & 90.50 & \framebox{93.35}\\
		L1\_OUt & 72.54 & \framebox{74.92} & 74.10 & \framebox{74.56} & 74.26 & 73.33\\
		L14\_B & 75.89 & \framebox{77.31} & \framebox{77.17} & 76.72 & 76.75 & 75.33\\
		L9\_sB & 84.73 & \framebox{86.01} & \framebox{85.84} & 85.40 & 85.78 & 82.27\\
		G1b & 79.74 & 75.94 & \framebox{80.73} & \framebox{81.66} & 77.67 & 79.49\\
		G3 & 76.22 & 64.54 & \framebox{78.84} & \framebox{78.78} & 72.67 & 71.24\\
		G6 & 83.13 & 83.50 & \framebox{84.38} & 83.96 & \framebox{84.28} & 74.90\\
		M1 & 79.12 & 73.09 & \framebox{80.93} & \framebox{81.89} & 76.77 & 77.61\\
		M4 & 64.73 & \framebox{68.04} & \framebox{67.91} & 67.48 & 67.89 & 61.36\\
		M6 & 81.25 & 80.09 & \framebox{83.06} & 82.45 & \framebox{83.21} & 78.02\\
		\noalign{\smallskip}\hline 
	\end{tabular} 
\end{center}   
\noindent\footnotesize Table 16.- Average  SVM accuracy (proportion of correct classification) outputs, over 200 runs of the considered methods with sample size $n=50$. \normalsize

\

\

\begin{center}\footnotesize
	\begin{tabular}{lcccccc}
		\multicolumn{7}{c}{\bf \small SVM number of variables}\\
		\hline\noalign{\smallskip}
		Model & MID & FCD & RD & VD & CD & Base\\
		\hline\noalign{\smallskip}
		L7\_OU & 12.1 & 14.1 & \framebox{10.1} & 11.5 & \framebox{11.1} & 100\\
		L1\_OUt & 8.9 & 8.0 & 7.0 & \framebox{5.9} & \framebox{6.6} & 100\\
		L14\_B & 7.9 & 8.0 & \framebox{6.0} & 7.5 & \framebox{5.9} & 100\\
		L9\_sB & \framebox{4.9} & 7.5 & \framebox{4.7} & 4.9 & 5.3 & 100\\
		G1b & 6.9 & 13.4 & \framebox{4.8} & \framebox{1.5} & 9.0 & 100\\
		G3 & 6.4 & 11.6 & \framebox{3.7} & \framebox{3.7} & 8.2 & 100\\
		G6 & 3.4 & \framebox{1.9} & 3.5 & 3.1 & \framebox{2.7} & 100\\
		M1 & 4.9 & 11.7 & \framebox{4.5} & \framebox{1.4} & 8.4 & 100\\
		M4 & 6.5 & 5.0 & 5.1 & \framebox{3.4} & \framebox{4.6} & 100\\
		M6 & 7.1 & 7.6 & \framebox{5.5} & \framebox{5.6} & 6.4 & 100\\
		\noalign{\smallskip}\hline 
	\end{tabular} 
\end{center}   
\noindent\footnotesize Table 17.- Average number of selected variables over 200 runs of the considered methods with sample size $n=50$ using  SVM.\normalsize

\section*{Results with quotient criterion and other classifiers}

The outputs included in the paper  correspond to the difference criterion (in order to combine the relevance and redundancy measures). We provide here some additional results for the quotient criterion, instead of the difference one. Besides, Figure 2 in the paper was produced using only the outputs for the $k$-NN classfier. In this document we show the analogous displays for LDA, SVM and NB.
Again, the entire simulation outputs can be downloaded from \url{www.uam.es/antonio.cuevas/exp/mRMR-outputs.xlsx}.

Tables 1a-9a  below correspond and Tables 1-9 in the main paper with the difference criterion replaced with the quotient one. Figures 2a, 2b and 2c correspond to Figure 2 using the other classifiers.

\begin{center}\footnotesize
	\begin{tabular}{llcccccc}
		\hline\noalign{\smallskip}
		\small Output (NB) & \small Sample size & \small MIQ & \small FCQ & \small RQ & \small VQ & \small CQ & \small Base\\
		\hline\noalign{\smallskip}
		\small Avgerage accuracy & $n=30$ & 78.10 & 78.76 & \framebox{79.53} & \framebox{79.58} & 79.12 & 77.28\\
		& $n=50$ & 79.59 & 79.73 & \framebox{80.86} & \framebox{80.81} & 80.26 & 78.29\\
		& $n=100$ & 80.62 & 80.54 & \framebox{81.82} & \framebox{81.75} & 81.16 & 78.84\\
		& $n=200$ & 81.24 & 81.03 & \framebox{82.48} & \framebox{82.35} & 81.77 & 79.13\\
		\noalign{\smallskip} \hline\noalign{\smallskip} 
		\small Average dim. red & $n=30$ & 8.9 & 8.6 & \framebox{7.2} & \framebox{7.0} & 7.9 & 100\\
		& $n=50$ & 8.3 & 8.1 & \framebox{6.7} & \framebox{6.7} & 7.4 & 100\\
		& $n=100$ & 7.7 & 7.2 & \framebox{6.1} & \framebox{6.1} & 6.9 & 100\\
		& $n=200$ & 7.1 & 6.7 & \framebox{5.7} & \framebox{5.9} & 6.5 & 100\\
		\noalign{\smallskip} \hline\noalign{\smallskip} 
		\small Victories over Base & $n=30$ & 60 & 65 & \framebox{72} & \framebox{76} & 68 & -\\
		& $n=50$ & 67 & 61 & \framebox{79} & \framebox{78} & 68 & -\\
		& $n=100$ & 71 & 64 & \framebox{88} & \framebox{86} & 79 & -\\
		& $n=200$ & 75 & 68 & \framebox{92} & \framebox{91} & 84 & -\\
		\noalign{\smallskip} \hline 
	\end{tabular} 
\end{center}   
\noindent\footnotesize Table 1a.- Performance outputs for the  considered methods, using NB and the quotient criterion, with different sample sizes. Each output is the result  of the 100 different models for each sample size.\normalsize    

\

\begin{center}\footnotesize
	\begin{tabular}{llcccccc}
		\hline\noalign{\smallskip}
		%&&\multicolumn{6}{c}{\small Methods} \\ \cline{3-8}\noalign{\smallskip}
		\small Output ($k$-NN) & \small Sample size & \small MIQ & \small FCQ & \small RQ & \small VQ & \small CQ & \small Base\\ 
		\hline\noalign{\smallskip}
		\small Avgerage accuracy & $n=30$ & 80.02 & 79.65 & \framebox{80.85} & \framebox{80.82} & 80.09 & 78.98\\
		& $n=50$ & 81.32 & 80.40 & \framebox{81.72} & \framebox{81.67} & 80.87 & 80.34\\
		& $n=100$ & \framebox{82.84} & 81.34 & \framebox{82.73} & 82.65 & 81.83 & 81.99\\
		& $n=200$ & \framebox{84.06} & 82.09 & \framebox{83.56} & 83.49 & 82.65 & 83.38\\
		\noalign{\smallskip} \hline \noalign{\smallskip}
		\small Average dim. red & $n=30$ & 9.3 & 9.5 & \framebox{7.4} & \framebox{7.6} & 8.1 & 100\\
		& $n=50$ & 9.6 & 9.6 & \framebox{7.6} & \framebox{7.8} & 8.3 & 100\\
		& $n=100$ & 9.9 & 9.9 & \framebox{8.0} & \framebox{8.2} & 8.6 & 100\\
		& $n=200$ & 10.1 & 10.1 & \framebox{8.3} & \framebox{8.5} & 9.0 & 100\\
		\noalign{\smallskip} \hline \noalign{\smallskip}
		\small Victories over Base & $n=30$ & 71 & 58 & \framebox{76} & \framebox{74} & 67 & -\\
		& $n=50$ & 67 & 53 & \framebox{73} & \framebox{72} & 64 & -\\
		& $n=100$ & \framebox{71} & 49 & \framebox{64} & \framebox{64} & 55 & -\\
		& $n=200$ & \framebox{64} & 42 & 62 & \framebox{64} & 54 & -\\
		\noalign{\smallskip}\hline  
	\end{tabular} 
\end{center}   
\noindent\footnotesize Table 2a.- Performance outputs for the  considered methods, using $k$-NN and the quotient criterion, with different sample sizes. Each output is the result  of the 100 different models for each sample size.\normalsize    

\

\begin{center}\footnotesize
	\begin{tabular}{llcccccc}
		\hline\noalign{\smallskip}
		\small Output (LDA) & \small Sample size & \small MIQ & \small FCQ & \small RQ & \small VQ & \small CQ & \small Base\\
		\hline\noalign{\smallskip}
		\small Avgerage accuracy & $n=30$ & \framebox{78.67} & 77.58 & \framebox{78.64} & \framebox{78.64} & 78.15 & 60.80\\
		& $n=50$ & \framebox{80.20} & 78.53 & \framebox{79.58} & 79.52 & 79.11 & 58.75\\
		& $n=100$ & \framebox{81.74} & 79.62 & \framebox{80.65} & 80.52 & 80.25 & 53.11\\
		& $n=200$ & \framebox{82.90} & 80.47 & \framebox{81.53} & 81.35 & 81.10 & 73.25\\
		\noalign{\smallskip}\hline\noalign{\smallskip}
		\small Average dim. red & $n=30$ & 5.8 & \framebox{4.7} & \framebox{4.7} & \framebox{4.7} & 5.1 & 100\\
		& $n=50$ & 6.9 & 5.7 & \framebox{5.6} & \framebox{5.6} & 6.0 & 100\\
		& $n=100$ & 8.3 & 7.1 & \framebox{6.9} & \framebox{7.0} & 7.3 & 100\\
		& $n=200$ & 9.5 & 8.3 & \framebox{8.0} & \framebox{8.1} & 8.4 & 100\\
		\noalign{\smallskip}\hline  
	\end{tabular} 
\end{center}   
\noindent\footnotesize Table 3a.- Performance outputs for the  considered methods, using LDA and the quotient criterion, with different sample sizes. Each output is the result  of the 100 different models for each sample size.\normalsize    

\

\begin{center}\footnotesize
	\begin{tabular}{llcccccc}
		\hline\noalign{\smallskip}
		\small Output (SVM) & \small Sample size & \small MIQ & \small FCQ & \small RQ & \small VQ & \small CQ & \small Base\\
		\hline\noalign{\smallskip}
		\small Avgerage accuracy & $n=30$ & \framebox{81.62} & 79.81 & 80.69 & 80.65 & 80.27 & \framebox{81.91}\\
		& $n=50$ & \framebox{82.69} & 80.42 & 81.43 & 81.35 & 80.96 & \framebox{82.99}\\
		& $n=100$ & \framebox{83.80} & 81.21 & 82.20 & 82.12 & 81.76 & \framebox{84.11}\\
		& $n=200$ & \framebox{84.61} & 81.79 & 82.90 & 82.76 & 82.42 & \framebox{84.91}\\
		\noalign{\smallskip}\hline\noalign{\smallskip}
		\small Average dim. red & $n=30$ & 10.6 & 10.3 & \framebox{9.1} & \framebox{9.2} & 9.5 & 100\\
		& $n=50$ & 10.7 & 10.4 & \framebox{9.3} & \framebox{9.4} & 9.7 & 100\\
		& $n=100$ & 11.1 & 10.5 & \framebox{9.5} & \framebox{9.7} & 9.9 & 100\\
		& $n=200$ & 11.4 & 10.7 & \framebox{9.7} & \framebox{9.9} & 10.0 & 100\\
		\noalign{\smallskip}\hline\noalign{\smallskip}
		\small Victories over Base & $n=30$ & 32 & 37 & \framebox{49} & \framebox{47} & 42 & -\\
		& $n=50$ & 35 & 34 & \framebox{51} & \framebox{52} & 44 & -\\
		& $n=100$ & 35 & 33 & \framebox{51} & \framebox{50} & 48 & -\\
		& $n=200$ & 33 & 31 & \framebox{52} & \framebox{51} & 48 & -\\
		\noalign{\smallskip}\hline  
	\end{tabular} 
\end{center}   
\noindent\footnotesize Table 4a.- Performance outputs for the  considered methods, using SVM and the quotient criterion, with different sample sizes. Each output is the result  of the 100 different models for each sample size.\normalsize    

\newpage

\begin{center} \footnotesize
	\begin{tabular}{llccccc}
		\hline\noalign{\smallskip}
		\small Ranking criterion (NB) & \small Sample size &\small MIQ & \small FCQ & \small RQ & \small VQ & \small CQ \\
		\hline\noalign{\smallskip}
		\small Relative & $n=30$ & 2.27 & 5.67 & \framebox{8.69} & \framebox{8.63} & 7.65\\
		& $n=50$ & 2.69 & 4.94 & \framebox{9.09} & \framebox{8.70} & 7.51\\
		& $n=100$ & 2.75 & 4.75 & \framebox{9.21} & \framebox{8.80} & 7.44\\
		& $n=200$ & 2.71 & 4.57 & \framebox{8.87} & \framebox{8.31} & 7.41\\
		\noalign{\smallskip}\hline\noalign{\smallskip}
		\small Positional & $n=30$ & 6.78 & 7.83 & \framebox{8.53} & \framebox{8.50} & 8.39\\
		& $n=50$ & 6.79 & 7.57 & \framebox{8.93} & \framebox{8.51} & 8.22\\
		& $n=100$ & 6.80 & 7.47 & \framebox{9.01} & \framebox{8.58} & 8.14\\
		& $n=200$ & 6.84 & 7.56 & \framebox{8.92} & \framebox{8.43} & 8.25\\
		\noalign{\smallskip}\hline\noalign{\smallskip}
		\small F1 & $n=30$ & 12.25 & 15.85 & \framebox{17.39} & \framebox{17.58} & 17.01\\
		& $n=50$ & 12.24 & 14.90 & \framebox{19.19} & \framebox{17.37} & 16.35\\
		& $n=100$ & 12.35 & 14.60 & \framebox{19.67} & \framebox{17.38} & 16\\
		& $n=200$ & 12.49 & 14.92 & \framebox{19.28} & \framebox{16.86} & 16.45\\
		\noalign{\smallskip}\hline     
	\end{tabular}   
\end{center} 
\footnotesize Table 5a.- Global scores of the considered (quotient-based) methods using three different ranking criteria with the NB classifier.\normalsize    

\

\begin{center} \footnotesize
	\begin{tabular}{llccccc}
		\hline\noalign{\smallskip}
		\small Ranking criterion ($k$-NN) & \small Sample size &\small MIQ & \small FCQ & \small RQ & \small VQ & \small CQ \\
		\hline\noalign{\smallskip}
		\small Relative & $n=30$ & 3.70 & 3.97 & \framebox{8.51} & \framebox{8.09} & 6.03\\
		& $n=50$ & 4.39 & 3.72 & \framebox{7.84} & \framebox{7.59} & 5.61\\
		& $n=100$ & 4.93 & 3.46 & \framebox{7.24} & \framebox{6.91} & 5.15\\
		& $n=200$ & 5.52 & 3.00 & \framebox{6.72} & \framebox{6.52} & 5.00\\
		\noalign{\smallskip}\hline\noalign{\smallskip}
		\small Positional & $n=30$ & 7.31 & 7.29 & \framebox{9.06} & \framebox{8.64} & 7.70\\
		& $n=50$ & 7.55 & 7.30 & \framebox{8.90} & \framebox{8.67} & 7.58\\
		& $n=100$ & 7.75 & 7.37 & \framebox{8.82} & \framebox{8.62} & 7.49\\
		& $n=200$ & 7.96 & 7.43 & \framebox{8.56} & \framebox{8.45} & 7.60\\
		\noalign{\smallskip}\hline\noalign{\smallskip}
		\small F1 & $n=30$ & 14.32 & 13.96 & \framebox{19.66} & \framebox{17.80} & 14.26\\
		& $n=50$ & 15.27 & 14.06 & \framebox{18.78} & \framebox{17.94} & 13.95\\
		& $n=100$ & 16.02 & 14.19 & \framebox{18.44} & \framebox{17.90} & 13.62\\
		& $n=200$ & 16.84 & 14.09 & \framebox{17.48} & \framebox{17.57} & 14.02\\
		\noalign{\smallskip}\hline      
	\end{tabular}   
\end{center} 
\footnotesize Table 6a.- Global scores of the considered (quotient-based) methods using three different ranking criteria with the $k$-NN classifier.\normalsize    

\

\begin{center} \footnotesize
	\begin{tabular}{llccccc}
		\hline\noalign{\smallskip}
		\small Ranking criterion (LDA) & \small Sample size &\small MIQ & \small FCQ & \small RQ & \small VQ & \small CQ \\
		\hline\noalign{\smallskip}
		\small Relative & $n=30$ & 3.94 & 3.14 & \framebox{7.48} & \framebox{7.38} & 5.38\\
		& $n=50$ & 4.26 & 2.86 & \framebox{6.97} & \framebox{6.66} & 5.19\\
		& $n=100$ & 4.76 & 2.60 & \framebox{6.98} & \framebox{6.43} & 5.33\\
		& $n=200$ & 5.27 & 2.36 & \framebox{6.78} & \framebox{6.13} & 5.23\\
		\noalign{\smallskip}\hline\noalign{\smallskip}
		\small Positional & $n=30$ & 7.49 & 6.99 & \framebox{9.01} & \framebox{8.96} & 7.55\\
		& $n=50$ & 7.64 & 7.12 & \framebox{8.90} & \framebox{8.64} & 7.70\\
		& $n=100$ & 7.72 & 7.13 & \framebox{8.89} & \framebox{8.52} & 7.74\\
		& $n=200$ & 7.80 & 7.23 & \framebox{8.79} & \framebox{8.36} & 7.82\\
		\noalign{\smallskip}\hline\noalign{\smallskip}
		\small F1 & $n=30$ & 15.05 & 12.67 & \framebox{19.11} & \framebox{19.32} & 13.85\\
		& $n=50$ & 15.63 & 12.95 & \framebox{18.91} & \framebox{18.07} & 14.44\\
		& $n=100$ & 15.80 & 13.04 & \framebox{18.83} & \framebox{17.72} & 14.61\\
		& $n=200$ & 16.25 & 13.29 & \framebox{18.62} & \framebox{17.04} & 14.80\\
		\noalign{\smallskip}\hline    
	\end{tabular}   
\end{center} 
\footnotesize Table 7a.- Global scores of the considered (quotient-based) methods using three different ranking criteria with LDA.\normalsize    

\

\begin{center} \footnotesize
	\begin{tabular}{llccccc}
		\hline\noalign{\smallskip}
		\small Ranking criterion (SVM)& \small Sample size &\small MIQ & \small FCQ & \small RQ & \small VQ & \small CQ \\
		\hline\noalign{\smallskip}
		\small Relative & $n=30$ & 6.02 & 2.85 & \framebox{6.58} & \framebox{6.28} & 4.42\\
		& $n=50$ & 5.99 & 2.72 & \framebox{6.70} & \framebox{6.15} & 4.73\\
		& $n=100$ & \framebox{6.14} & 2.61 & \framebox{6.43} & 5.99 & 4.66\\
		& $n=200$ & \framebox{6.42} & 2.30 & \framebox{6.29} & 5.75 & 4.68\\
		\noalign{\smallskip}\hline\noalign{\smallskip}
		\small Positional & $n=30$ & 8.26 & 7.20 & \framebox{8.74} & \framebox{8.46} & 7.34\\
		& $n=50$ & 8.16 & 7.19 & \framebox{8.80} & \framebox{8.43} & 7.42\\
		& $n=100$ & 8.26 & 7.31 & \framebox{8.66} & \framebox{8.32} & 7.48\\
		& $n=200$ & \framebox{8.28} & 7.36 & \framebox{8.61} & 8.22 & 7.56\\
		\noalign{\smallskip} \hline\noalign{\smallskip}
		\small F1 & $n=30$ & \framebox{17.90} & 13.78 & \framebox{17.99} & 17.17 & 13.16\\
		& $n=50$ & \framebox{17.58} & 13.51 & \framebox{18.21} & 17.28 & 13.42\\
		& $n=100$ & \framebox{17.97} & 13.86 & \framebox{17.71} & 16.95 & 13.59\\
		& $n=200$ & \framebox{17.89} & 13.85 & \framebox{17.70} & 16.58 & 14.06\\
		\noalign{\smallskip}\hline      
	\end{tabular}   
\end{center} 
\footnotesize Table 8a.- Global scores of the considered (quotient-based) methods using three different ranking criteria with the linear SVM.\normalsize    

\

\begin{center}\footnotesize
	\begin{tabular}{llcccccc}   
		\multicolumn{8}{c}{\small \bf NB outputs}\\
		\hline\noalign{\smallskip}
		\small Output & \small Data & \small MIQ & \small FCQ & \small RQ & \small VQ & \small CQ & \small Base\\
		\hline\noalign{\smallskip} 
		\small Classification accuracy & Growth & \framebox{88.17} & \framebox{87.10} & 86.02 & 86.02 & 87.10 & 84.95\\
		& Tecator & 96.28 & 97.67 & \framebox{99.53} & \framebox{99.53} & 98.14 & 97.21\\
		& Phoneme & 73.08 & 80.20 & \framebox{80.38} & 80.15 & \framebox{80.32} & 74.08\\
		\noalign{\smallskip} \hline\noalign{\smallskip} 
		\small Number of variables & Growth & 1.5 & \framebox{1.1} & \framebox{1.1} & \framebox{1.1} & \framebox{1.1} & 31\\
		& Tecator & 4.8 & 5.0 & \framebox{1.0} & \framebox{1.0} & 4.4 & 100\\
		& Phoneme & 14.5 & \framebox{10.6} & 16.7 & 16.8 & \framebox{14.1} & 256\\
		\noalign{\smallskip} \hline
		\\  
		\multicolumn{8}{c}{\small \bf $k$-NN outputs}\\
		\hline\noalign{\smallskip}
		\small Output & \small Data & \small MIQ & \small FCQ & \small RQ & \small VQ & \small CQ & \small Base\\
		\hline\noalign{\smallskip} 
		\small Classification accuracy & Growth & \framebox{95.70} & 83.87 & 83.87 & 83.87 & 83.87 & \framebox{96.77}\\
		& Tecator & 96.74 & 99.07 & \framebox{99.53} & \framebox{99.53} & 98.60 & 98.60\\
		& Phoneme & 75.53 & \framebox{81.42} & 79.79 & 80.38 & \framebox{80.61} & 78.80\\
		\noalign{\smallskip} \hline\noalign{\smallskip} 
		\small Number of variables & Growth & 3.9 & \framebox{1.0} & \framebox{1.0} & \framebox{1.0} & \framebox{1.0} & 31\\
		& Tecator & 4.0 & 3.0 & \framebox{1.0} & \framebox{1.0} & 4.3 & 100\\
		& Phoneme & 18.4 & \framebox{12.1} & 12.3 & 15.2 & \framebox{6.7} & 256\\
		\noalign{\smallskip} \hline
		\\
		\multicolumn{8}{c}{\small \bf LDA outputs}\\
		\hline\noalign{\smallskip}
		\small Output & \small Data & \small MIQ & \small FCQ & \small RQ & \small VQ & \small CQ & \small Base\\
		\hline\noalign{\smallskip} 
		\small Classification accuracy & Growth & \framebox{95.70} & \framebox{91.40} & \framebox{91.40} & \framebox{91.40} & \framebox{91.40} & -\\
		& Tecator & \framebox{94.88} & 94.42 & \framebox{94.88} & 94.42 & \framebox{95.35} & -\\
		& Phoneme & 74.55 & 78.88 & 79.10 & \framebox{79.63} & \framebox{80.26} & -\\
		\noalign{\smallskip} \hline\noalign{\smallskip} 
		\small Number of variables & Growth & \framebox{3.7} & 5.0 & \framebox{4.9} & \framebox{4.9} & 5.0 & -\\
		& Tecator & 6.1 & 8.4 & 4.1 & \framebox{2.2} & \framebox{3.1} & -\\
		& Phoneme & 19.0 & \framebox{8.9} & 10.0 & \framebox{9.0} & 9.2 & -\\
		\noalign{\smallskip} \hline
		\\
		\multicolumn{8}{c}{\small \bf SVM outputs}\\
		\hline\noalign{\smallskip}
		\small Output & \small Data & \small MIQ & \small FCQ & \small RQ & \small VQ & \small CQ & \small Base\\
		\hline\noalign{\smallskip} 
		\small Classification accuracy & Growth & \framebox{94.62} & 87.10 & 87.10 & 87.10 & 86.02 & \framebox{95.70}\\
		& Tecator & 98.14 & \framebox{99.07} & \framebox{99.53} & \framebox{99.07} & 98.60 & \framebox{99.07}\\
		& Phoneme & 75.30 & \framebox{80.71} & 80.67 & 80.37 & 80.33 & \framebox{80.96}\\
		\noalign{\smallskip} \hline\noalign{\smallskip} 
		\small Number of variables & Growth & \framebox{3.5} & 5.0 & \framebox{4.9} & \framebox{4.9} & 5.0 & 31\\
		& Tecator & 6.7 & 2.1 & \framebox{1.0} & \framebox{1.0} & 4.1 & 100\\
		& Phoneme & 19.3 & \framebox{10.1} & 11.3 & \framebox{10.8} & 12.2 & 256\\
		\noalign{\smallskip} \hline
	\end{tabular} 
\end{center}
{\footnotesize Table 9a.- Performances of the different (quotient-based) mRMR methods in three real data sets. From top to bottom tables stand for Naive Bayes, $k$-NN, LDA and linear SVM outputs respectively.}

\

\begin{center}
	\includegraphics[height=5.5cm,width=15cm]{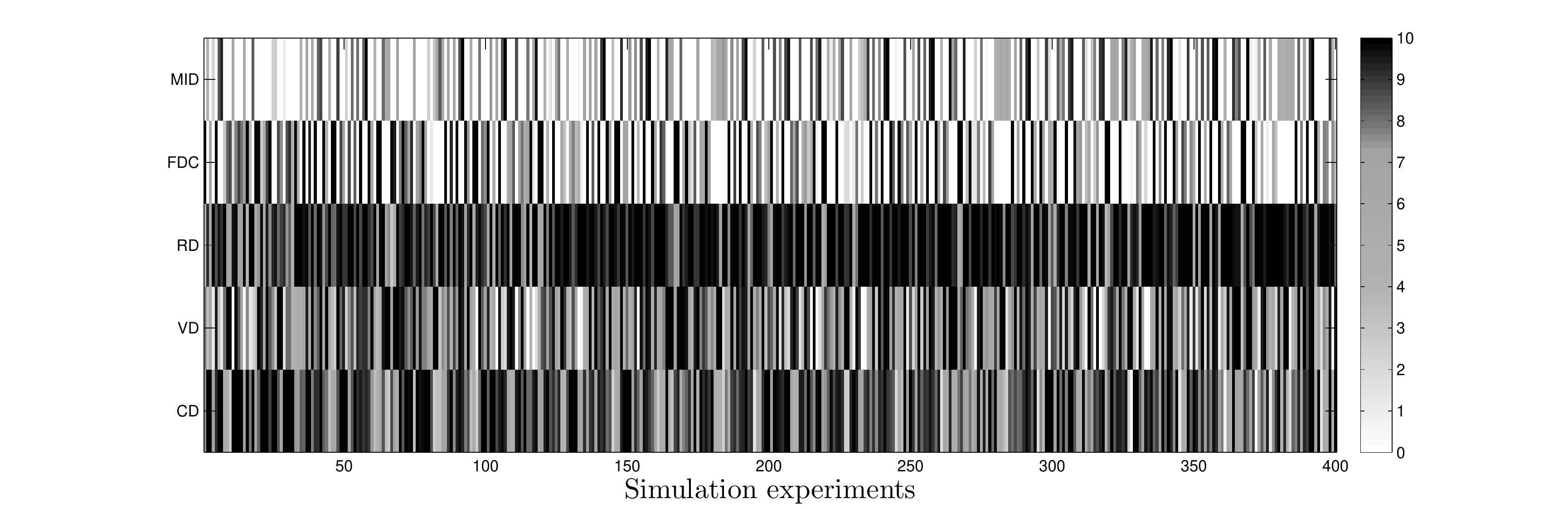}
\end{center}
{\footnotesize Figure 2a.- Cromatic version of the global relative ranking table taking into account the 400 considered experiments (columns) and the Naive Bayes classifier. }

\begin{center}
	\includegraphics[height=5.5cm,width=15cm]{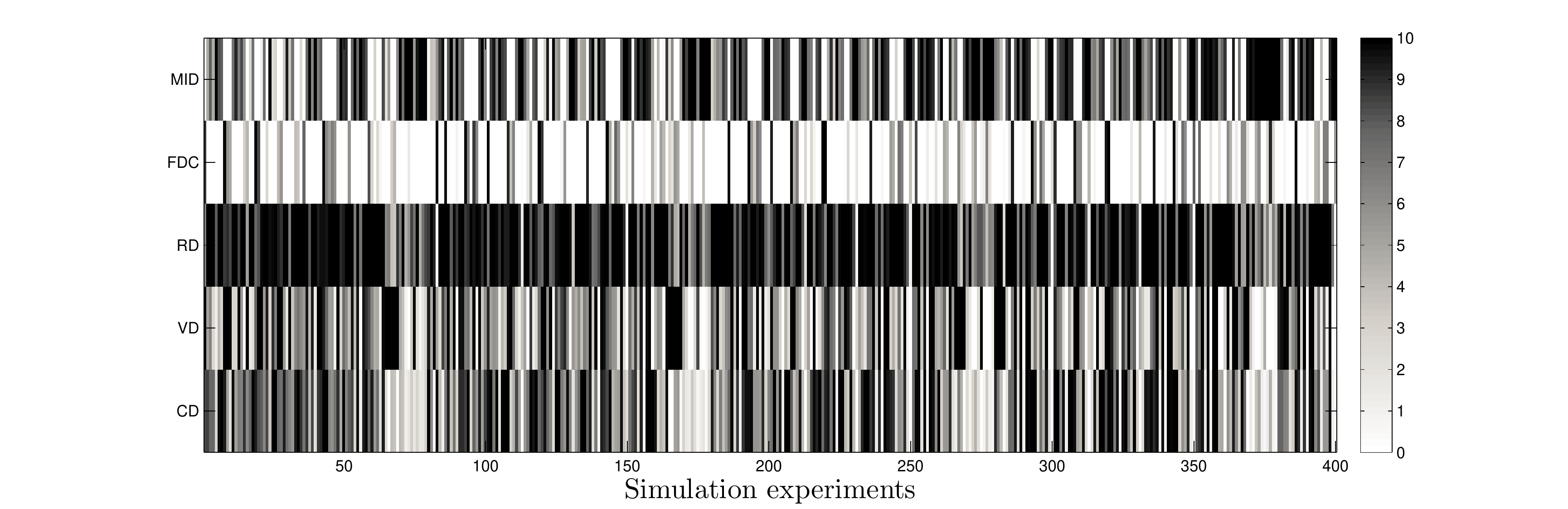}
\end{center}
{\footnotesize Figure 2b.- Cromatic version of the global relative ranking table taking into account the 400 considered experiments (columns) and the Linear Discriminant Analysis. } 

\begin{center}
	\includegraphics[height=5.5cm,width=15cm]{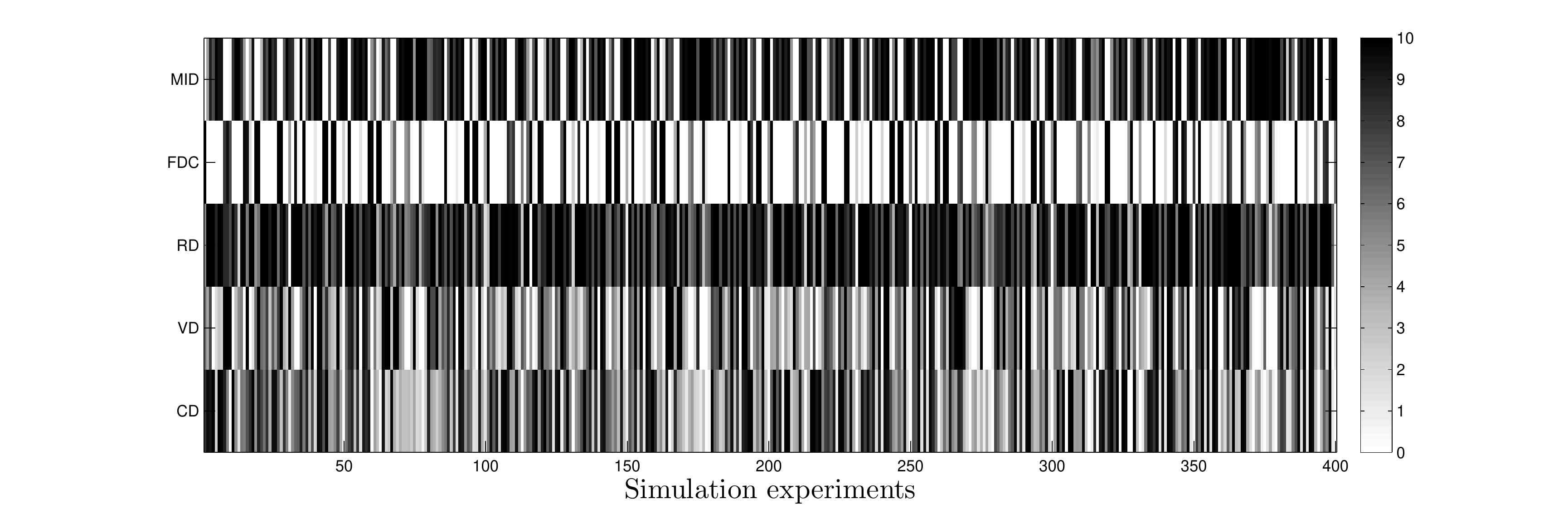}
\end{center}
{\footnotesize Figure 2c.- Cromatic version of the global relative ranking table taking into account the 400 considered experiments (columns) and the linear SVM. } 


\begin{thebibliography}{99}
										
										
										\bibitem[\protect\citeauthoryear{Golub \it et al\rm.}{1999}]{gol99}
										\rm{Golub, T.R. and Slonim, D.K. and Tamayo, P. and Huard, C. and Gaasenbeek, M. and Mesirov, J.P. and Coller, H. and Loh, M.L. and Downing, J.R. and Caligiuri, M.A. and others}.
										Molecular classification of cancer: class discovery and class prediction by gene expression monitoring.
										\textit{Science}.1999;286:531--537.
										
										
										\bibitem[\protect\citeauthoryear{Lindquist and  McKeague}{2009}]{lin09}
										\rm{Lindquist, M.A. and McKeague, I.W.} 
										\newblock Logistic regression with Brownian-like predictors.
										\textit{Journal of the American Statistical Association}. 2009;104:1575–-1585.
										
										\bibitem[\protect\citeauthoryear{Guyon  \it et al\rm.}{2006}]{guy06}
										\rm{Guyon, I. and Gunn, S. and Nikravesh, M. and Zadeh, L.A}. 
										\newblock Feature Extraction: Foundations and Applications\rm.
										Springer. 2006.
										
										\bibitem[\protect\citeauthoryear{Berrendero \it et al\rm.}{2014}]{ber14}
										\rm{Berrendero, J.R., Cuevas, A. and Torrecilla, J.L}.
										\newblock Variable selection in functional data classification: a maxima hunting proposal.
										\it Unpublished manuscript\rm. 2014.
										
										
										\bibitem[\protect\citeauthoryear{Ba\'{\i}llo \it et al\rm.}{2011b}]{bai11b}
										\rm{Ba\'{\i}llo, A., Cuevas, A. and Fraiman, R}. 
										Classification methods with functional data. In \it Oxford Handbook of Functional Data Analysis\rm, 2011;pp-259--297. In: Ferraty F. and Romain Y., editors. Oxford University Press.
										
										
										
										\bibitem[\protect\citeauthoryear{Ba\'{\i}llo \it et al\rm.}{2011a}]{bai11a}
										\rm{Ba\'{\i}llo, A., Cuesta-Albertos, J. A. and Cuevas, A}. 
										Supervised classification for a family of Gaussian functional models.
										\textit{Scand. J. Stat}. 2011;38:480--498.
										
										
										\bibitem[\protect\citeauthoryear{Ding and Peng}{2005}]{din05}
										\rm{Ding, C. and Peng, H}. 
										Minimum redundancy feature selection from microarray gene
										expression data. 
										{J. Bioinform. Comput. Biol.} 2005;3:185--205.
										
										
										\bibitem[\protect\citeauthoryear{Peng \it et al\rm.}{2005}]{pen05}
										\rm{Peng, H., Long, F. and Ding, C.} 
										\newblock Feature selection based on mutual information: criteria
										of max-dependency, max-relevance, and min-redundancy. 
										{IEEE Trans. Pattern Anal. Mach. Intell.} 2005;27:1226--1238.
										
										
										
										\bibitem[\protect\citeauthoryear{Battiti}{1994}]{bat94}
										\rm{Battiti, R}. 
										Using mutual information for selecting features in supervised neural net learning.
										\textit{Neural Networks, IEEE Transactions on}. 1994;5:537--550.
										
										
										\bibitem[\protect\citeauthoryear{Kwak and Choi}{2002}]{kwa02}
										\rm{Kwak, N. and Choi, C.H.}
										\newblock Input feature selection by mutual information based on Parzen window.
										\textit{Pattern Analysis and Machine Intelligence, IEEE Transactions on}. 2002;24:1667--1671
										
										
										
										
										\bibitem[\protect\citeauthoryear{Yu and Liu}{2004}]{yu04}
										\rm{Yu, L. and Liu, H.} 
										Efficient feature selection via analysis of relevance and redundancy.
										\textit{The Journal of Machine Learning Research}. 2004;5:1205--1224.
										
										
										\bibitem[\protect\citeauthoryear{Sz\'{e}kely \it et al\rm.}{2007}]{sze07}
										\rm{Sz\'{e}kely, G. J., Rizzo, M. L. and Bakirov, N. K.} 
										Measuring and testing dependence by correlation of distances.
										\textit{Ann. Statist.} 2007;35:2769--2794.
										
										
										
										\bibitem[\protect\citeauthoryear{Sz\'{e}kely and Rizzo}{2009}]{sze09}
										\rm{Sz\'{e}kely, G. J. and Rizzo, M. L.} 
										Brownian Distance Covariance.
										\textit{Ann. Appl. Stat.} 2009;3:1236--1265.
										
										\bibitem[\protect\citeauthoryear{Hall and Miller}{2011}]{hal11}
										\rm{Hall, P. and Miller, H.}
										Determining and depicting relationships
										among components in high-dimensional
										variable selection.
										\textit{J. Comput. Graph. Statist}. 2011;20:988--1006.
										
										\bibitem[\protect\citeauthoryear{Reshef \it et al\rm.}{2011}]{res11}
										\rm{Reshef, D. N., Reshef, Y. A.,  Finucane, H. K.,  Grossman, S. R.
											McVean, G.,  Turnbaugh, P. J.,  Lander, E. S.,
											Mitzenmacher, M. and Sabeti, P. C.} 
										Detecting novel associations in large data sets.
										\textit{Science}\/ 2011;334:1518--1524.
										
										
										\bibitem[\protect\citeauthoryear{Sz\'{e}kely and Rizzo}{2012}]{sze12}
										\rm{Sz\'{e}kely, G. J. and Rizzo, M. L.} 
										On the uniqueness of distance covariance.
										\textit{Statist. Probab. Lett}. 2012;82:2278--2282.
										
										\bibitem[\protect\citeauthoryear{Sz\'{e}kely and Rizzo}{2013}]{sze13}
										\rm{Sz\'{e}kely, G. J. and Rizzo, M. L.} 
										Energy statistics: a class of statistics based on distances.
										\textit{J. Plann. Statist. Infer}. 2013;143:1249--1272.
										
										
										
										\bibitem[\protect\citeauthoryear{Wand and Jones}{1995}]{wan95}
										\rm{Wand, M.P. and Jones, M.C.} 
										\newblock \it Kernel smoothing\rm.
										Chapman \& Hill. 1995.
										
										\bibitem[\protect\citeauthoryear{Cao \it et al\rm.}{1994}]{cao94}
										\rm{Cao, R. and Cuevas, A. and Gonzalez-Manteiga, W}.  
										A comparative study of several smoothing methods in density estimation
										\textit{Computational Statistics \& Data Analysis}. 1994;17:153--176.
										
										
										\bibitem[\protect\citeauthoryear{Est{\'e}vez  \it et al\rm.}{2009}]{est09}
										\rm{Est{\'e}vez, P.A. and Tesmer, M, and Perez, C.A. and Zurada, J.M}. 
										\newblock Normalized mutual information feature selection. 
										\textit{Neural Networks, IEEE Transactions on} 2009;20:189--201.
										
										
										
										\bibitem[\protect\citeauthoryear{Mandal and  Mukhopadhyay}{2014}]{man14}
										\rm{Mandal, M. and n Mukhopadhyay, A.} 
										\newblock A novel PSO-based graph-theoretic approach for identifying most
										relevant and non-redundant gene markers from gene expression data.
										To appear in \textit{International Journal of Parallel, Emergent and Distributed Systems}. 2014.
										
										\bibitem[\protect\citeauthoryear{Nguyen \textit{et al.}}{2014}]{ngu14}
										\rm{Nguyen, X.V. and Chan, J. and Romano, S. and Bailey, J.} 
										\newblock Effective global approaches for mutual information based feature selection.
										\textit{Proceedings of the 20th ACM SIGKDD international conference on Knowledge discovery and data mining}. 2014:512--521.
										
										
										\bibitem[\protect\citeauthoryear{Demler \it et al\rm.}{2013}]{dem13}
										\rm{Demler, D.V., Pencina, M.J. and D'Agostino, R.B}. 
										\newblock Impact of correlation on predictive ability of biomarkers. 
										\textit{Stat Med.} 2013;32:4196--4210. 
										
										
										\bibitem[\protect\citeauthoryear{Hastie  \it et al\rm.}{2005}]{has05}
										\rm{Hastie, T. and Tibshirani, R. and Friedman, J. and Franklin, J}. 
										\newblock \it The elements of statistical learning: data mining, inference and prediction\rm.
										Springer. 2005.
										
										\bibitem[\protect\citeauthoryear{Hand}{2006}]{han06}
										\rm{Hand, D.}
										\newblock \it Classifier technology and the illusion of progress\rm.
										\it Statist. Sci\rm. 2006;21:1--34.
										
										
										
										\bibitem[\protect\citeauthoryear{Fan  \it et al\rm.}{2008}]{fan08}
										\rm{Fan, R.-E. and Chang, K.-W. and Hsieh, C.-J, and Wang, X.-R. and Lin C.-J.} 
										\newblock \it {LIBLINEAR}: A Library for Large Linear Classification\rm.
										\textit{Journal of Machine Learning Research} 2008;9:1871--1874.
										
										
										\bibitem[\protect\citeauthoryear{Ramsay and Silverman}{2005}]{ram05}
										\rm{Ramsay, J.O. and Silverman, B.W.} 
										\newblock \it Functional data analysis\rm.
										Springer. 2005.
										
										\bibitem[\protect\citeauthoryear{Ferraty and Vieu}{2006}]{fer06}
										\rm{Ferraty, F. and Vieu, P}. 
										\newblock \it Nonparametric Functional Data Analysis: Theory and Practice\rm.
										Springer. 2006.
										
										
										
										
										
										
										
									\end{thebibliography}
\end{document}